\documentclass[12pt, draftclsnofoot, onecolumn]{IEEEtran}
\usepackage{amsmath,epsfig,epsf,times,amssymb,booktabs,subfigure}
\usepackage{setspace}
\usepackage{array}
\usepackage{url}
\usepackage{float}
\usepackage{color}
\usepackage{graphicx}
\usepackage{cite}
\usepackage{enumerate}

\def\gap{0.75ex}
\abovedisplayskip\gap
\belowdisplayskip\gap
\abovedisplayshortskip\gap
\belowdisplayshortskip\gap

\graphicspath{{figs/}}

\newtheorem{proposition}{Proposition}

\newtheorem{corollary}{Corollary}
\newtheorem{remark}{Remark}

\newcommand{\remarkend}{ \IEEEQEDopen}
\newtheorem{lemma}{Lemma}

\allowdisplaybreaks

\begin{document}

\sloppy

\title{Explore and Eliminate: Optimized Two-Stage Search for Millimeter-Wave Beam Alignment}

\author{Min Li, Chunshan Liu, Stephen V. Hanly, Iain B. Collings, and Philip Whiting~\thanks{This work has been presented in part in Proceedings of the 53rd IEEE International Conference on Communications, ICC'19, Shanghai, China, May 20-24, 2019~\cite{min2019ICC}. Min Li is with College of Information Science and Electronic Engineering, Zhejiang University, Hangzhou, 310027, China (e-mail: min.li@zju.edu.cn). The other authors are with the School of Engineering, Macquarie University, Sydney, NSW 2109, Australia (e-mail: \{chunshan.liu, stephen.hanly, iain.collings, philip.whiting\}@mq.edu.au). (\textit{Corresponding author: Chunshan Liu})}}
\maketitle

\begin{abstract}
Swift and accurate alignment of transmitter (Tx) and receiver (Rx) beams is a fundamental design challenge to enable reliable outdoor millimeter-wave communications. In this paper, we propose a new Optimized Two-Stage Search (OTSS) algorithm for Tx-Rx beam alignment via spatial scanning. In contrast to one-shot exhaustive search, OTSS judiciously divides the training energy budget into two stages. In the first stage, OTSS explores and trains all candidate beam pairs and then eliminates a set of less favorable pairs learned from the received signal profile. In the second stage, OTSS takes an extra measurement for each of the survived pairs and combines with the previous measurement to determine the best one. For OTSS, we derive an upper bound on its misalignment probability, under a single-path channel model with training codebooks having an ideal beam pattern. We also characterize the decay rate function of the upper bound with respect to the training budget and further derive the optimal design parameters of OTSS that maximize the decay rate. OTSS is proved to asymptotically outperform state-of-the-art beam alignment algorithms, and  is numerically shown to achieve better performance with limited training budget and practically synthesized beams.
\end{abstract}


\section{Introduction}

\par Millimeter-wave (mmWave) communications has been recognized as one of the important technologies in the evolving 5G New Radio (NR)~\cite{pi2011introduction,andrews2014will,xiao2017millimeter,lee2018spectrum}. Owing to the abundant spectrum at mmWave bands (30-300 GHz), mmWave technology has great potential for enabling a variety of data-hungry mobile applications, such as video streaming and vehicle-to-vehicle communications~\cite{choi2016millimeter}. However, the unfavorable characteristics of mmWave bands, manifesting in severe path loss, sparse scattering and sensitivity to blockage, have posed great challenges in realizing reliable mmWave communications in practice~\cite{liu2018Mag}.

\par To combat the significant path loss, directional transmission via beamforming is necessary in particular for outdoor long-range mmWave communications. Establishment of such transmission, however, requires swift and accurate alignment of transmitter (Tx) and receiver (Rx) beams, which is non-trivial to accomplish. In this paper, we focus on this fundamental beam-alignment problem and advance existing studies by developing a new beam-alignment strategy.

\par Beam training via spatial scanning is a common approach for beam alignment in mmWave communications and it has drawn considerable attention from both academia and industry~\cite{liu2018Mag,wang2009beam,hur2013millimeter,alkhateeb2014channel,Xiao2016,liu2017Jsac,zhang2017codebook,haghighatshoar2016beam,kokshoorn2017millimeter,alkhateeb2017initial,va2018inverse,3GPPR1,3GPPR2,raghavan2018steady}. This approach involves a search through pre-defined beam codebooks that cover the scanning space to determine the best beam that aligns with the dominant path for communication. Depending on the application scenarios of mmWave communications, spatial scanning can be performed at one side of the communication link, e.g., at Tx or Rx, to find the best transmit/receive beam for data communication, or be performed at both sides of the communication link to find the best transmit and receive beam pair. In what follows, we discuss spatial scanning schemes by assuming that the search is two-sided. However, the algorithms and analysis will also apply to the one-sided scenario as a special case. Exhaustive search and hierarchical search are the two classic strategies of spatial scanning and they differ in both codebook construction and search mechanism~\cite{liu2018Mag}.

\par In exhaustive search, a training codebook is formed by narrow beams with large beamforming gain. Tx and Rx sequentially train each of the beam-pairs in the codebook and find the best one that maximizes a given performance metric (such as combined beamforming gain). On the other hand, in hierarchical search, multi-level codebooks are formed and arranged in a hierarchial manner by using fewer wider beams in the lower level and more narrower beams in the higher level to cover the same scanning space~\cite{wang2009beam,hur2013millimeter,alkhateeb2014channel,Xiao2016,liu2017Jsac,zhang2017codebook}. In the spirit of bisection search, Tx and Rx first train wide beam-pairs in a lower-level codebook and retain the best one, and then iteratively refine the search using the next-level codebook within the beam subspace associated with the survived beam-pair.

\par Compared with exhaustive search, hierarchical search reduces the search space, examines fewer beam-pairs and thus requires fewer measurements. However, this measurement reduction does not necessarily imply the superiority of hierarchical search. {In fact, if hierarchical search were to be performed at the highest possible beam switching speed (hence with the minimum time), the beam alignment performance would be very poor for low SNR users, owing to the misalignment error propagation originating from wide beams with small beamforming gain in an early stage~\cite{liu2017Jsac}.} To enhance its performance, a longer training time~\cite{liu2017Jsac} to accumulate more training energy is needed for each measurement to boost the effective SNR at receiver. With longer training time, exhaustive search becomes feasible. Compared with hierarchical search, exhaustive search examines more beam-pairs and thus takes more measurements; however, it enjoys high beamforming gain from the narrow beams adopted and might require less training energy per measurement to achieve a desired SNR at receiver. Therefore, it is unclear what the relative performance of these two search strategies is, subjected to the same \textit{total amount} of training energy budget. In~\cite{liu2017Jsac}, we have characterized the asymptotic misalignment probability of both strategies and proved that exhaustive search asymptotically outperforms hierarchical search when the training budget grows large, under the single-path channel model and with ideal beam codebooks.

\par In the meantime, the impact of practical beam codebook design on the performance of hierarchical search has also been studied and different beam synthesis techniques have been developed in~\cite{Xiao2016,zhang2017codebook,xiao2017codebook,xiao2018enhanced}. Variations of exhaustive search and hierarchical search have also been studied in the presence of favorable beam-pointing side information~\cite{va2018inverse} and in the context of multi-user scenarios~\cite{lin2017subarray}, respectively.

\par {The focus of our paper is outdoor mmWave systems where the pre-beamforming SNR is often much smaller than 0 dB. In this case, more training time is needed to ensure good beam alignment accuracy and we consider training times that are sufficient to perform exhaustive search. We propose a new search algorithm for beam alignment that uses the same training codebook as that of exhaustive search.} However, we compress the exhaustive search into a shorter time period (using the same beamforming vectors), and hence use less energy for that search. We then select a small set of promising beamforming directions, based on the initial search, and use the remaining time and energy to re-explore those directions, the other directions having been eliminated by the first stage. There is a particular payoff from this strategy in mmWave communications because the pre-beamforming SNR is low, and hence false positives can often occur in the standard exhaustive search, and also in the standard hierarchical search (particularly in the initial widebeam search). In our approach we are able to eliminate directions that have a low return in the first stage, and then focus the residual energy in exploring the remaining directions to eliminate the false positives. In the second stage, an extra measurement is taken and is coherently combined with the previous measurement for each of the remaining beam pairs. Among them, the algorithm recommends the one with the largest combined received energy as the final decision. We use the term ``Optimized Two-Stage Search (OTSS)'' to denote the algorithm.

\par Under a line-of-sight single-path channel model and with ideal beam patterns, we derive an upper bound on the misalignment probability of OTSS as a function of  key system parameters and verify its tightness numerically. Using large deviations techniques, we also characterize the decay rate function of the upper bound with respect to the total training energy budget, and further derive the optimal number of less favorable beam pairs eliminated in the first stage ($K^*$) and the optimal fraction of training budget allocated to the first stage ($\alpha^*$) that maximize the decay rate. Both $K^*$ and $\alpha^*$ depend only on the number of candidate beam pairs in the model. This analysis not only provides important guidance on the asymptotically optimal choice of the key design parameters of OTSS, but also allows us to conclude that OTSS asymptotically outperforms state-of-the-art baselines (including the classic hierarchical search and exhaustive search), when the same training energy budget for all strategies grows large. The performance advantage of OTSS is also verified numerically when the training budget is finite and when practically synthesized beams are adopted.

\par \textit{Notation}: Boldface uppercase and lowercase letters denote matrices and vectors, respectively, e.g., $\mathbf{A}$ is a matrix and $\mathbf{a}$ is a vector. $\|\mathbf{a}\|_2$ denotes the $l_2$ norm of vector $\mathbf{a}$. Notation $(\cdot)^T$ denotes the matrix transpose, while $(\cdot)^\dag$ denotes the conjugate transpose. For a pair of integers $(z_1, z_2)$ where $z_1 \le z_2$, $\left[z_1:z_2\right]$ is used to denote the discrete interval $\{z_1,z_1+1,\cdots,z_2\}$. Finally, ${\mathcal{CN}}(0,\sigma^2)$ denotes a complex Gaussian distribution with zero mean and variance $\sigma^2$.

\section{Beam Alignment Problem and Preliminaries}\label{sec:system:model}
\par We consider a point-to-point mmWave beam alignment problem, in which a Tx and a Rx wish to align their transmit/reception beams along the dominant path in a mmWave channel. We assume reliable feedback links (via, say, low frequencies) are available for the coordination of the beam search and that the Tx and the Rx are synchronized. We refer the readers to the literature for mmWave synchronization techniques~\cite{liu2017design,barati2014dreictional}.

\par
\begin{figure}[t]
\centering
\includegraphics[width=0.8\textwidth]{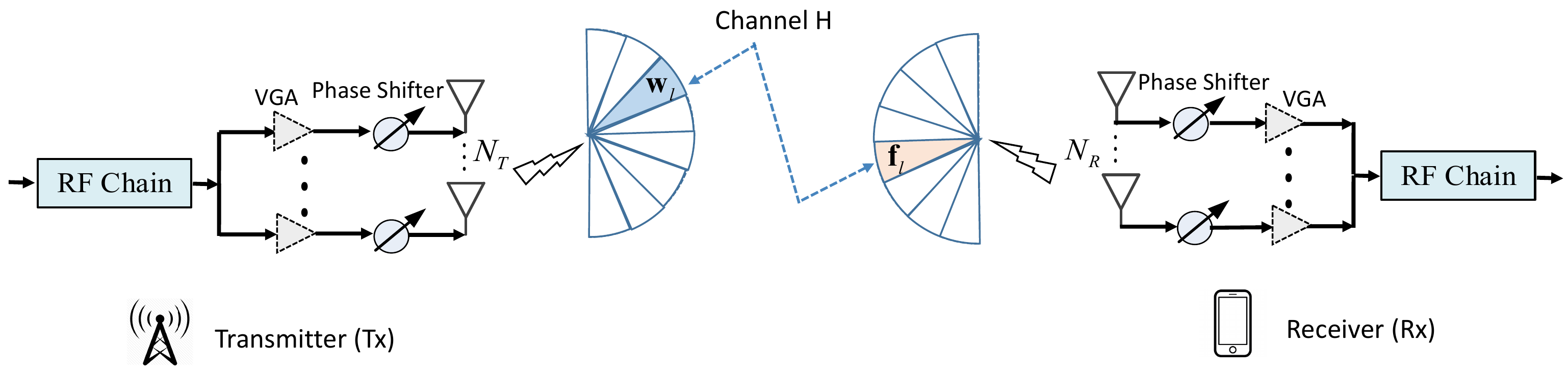}\caption{An illustration of the beam alignment problem. Tx and Rx wish to determine the best beam pair from a pre-defined Tx-Rx beam codebook that aligns with the dominant path in mmWave channel ${\mathbf H}$ (namely, that maximizes the effective channel gain $|{\bf f}_{{l}}^\dag {\mathbf H}{\bf w}_{{l}}|^2$).}
\label{fig:system}
\end{figure}

 In particular, beam training via spatial scanning is adopted as in~\cite{hur2013millimeter,liu2017Jsac}, see Fig.~\ref{fig:system} for an illustration. Specifically, let $\Psi$ and $\Phi$ be the entire Angle of Departure (AoD) and Angle of Arrival (AoA) scanning interval, respectively. Assume that Tx and Rx is equipped with $N_T$ and $N_R$ antennas, respectively. Let ${\mathcal C}_{T} =\{{\bar {\bf w}_{l_T}}\in {\mathbb C}^{N_T \times 1}, l_T \in [1:L_T]\}$ be a set of $L_T$ unit-norm beams at Tx that jointly cover the entire AoD and ${\mathcal C}_{R} =\{{\bar {\bf f}_{l_R}}\in {\mathbb C}^{N_R \times 1}, l_R \in [1:L_R]\}$ be a set of $L_R$ unit-norm beams at Rx that jointly cover the entire AoA. The Tx-Rx beam codebook ${\mathcal C}$ is therefore given by the cartesian product of ${\mathcal C}_{T}$ and ${\mathcal C}_{R}$ as ${\mathcal C}= \{({\bf w, \bf f}): {\bf w} \in {\mathcal C}_{T}, {\bf f} \in {\mathcal C}_{R}\}$, with $N \triangleq L_TL_R$ beam pairs in total. For ease of exposition, we simply use $({{\bf w}_l, {\bf f}_l})$ to denote the $l$th beam pair in ${\mathcal C}$, where $l \in [1:N]$. When constructing the Tx and Rx codebooks, we assume that both the magnitude and the phase of each beamforming coefficient are adjustable. Such codebooks can be realised when one variable gain amplifier (VGA) and one phase shifter is equipped for each antenna, as proposed in~\cite{Castellanos2018ICC,karacora2019hybrid} (see Fig.~\ref{fig:system} for an illustration), or be realised when two phase shifters are provided for each antenna~\cite{zhang2014achieving}.

\par Consider a frequency-flat and block-fading channel model, where the channel remains unchanged during the beam-alignment process. Let ${\mathbf H} \in {\mathbb C}^{{N_R}\times {N_T}}$ be an arbitrary realization of the mmWave channel between Tx and Rx. The goal of beam alignment is to determine the best beam pair $({{\bf w}_{{l}_{\text{opt}}}, {\bf f}_{{l}_{\text{opt}}}}) \in {\mathcal C}$ that maximizes the effective channel gain after beamforming, i.e.,
\begin{align}
l_{\text{opt}} = \mathop {\arg \max }\limits_{l \in [1:N]} |{\bf f}_{{l}}^\dag {\mathbf H}{\bf w}_{{l}}|^2. \label{equ:perfect:alignment}
\end{align}
However, since neither Tx nor Rx has knowledge of ${\mathbf H}$, it is necessary to carry out proper beam training by letting Tx transmit pilot symbols and Rx measure the pilots using beam pairs in ${\mathcal C}$. The Rx then selects the best beam pair based on the channel output measurements. In this work, we assume that there is a pre-defined total training time for spatial scanning and that the transmission power of the pilots is constant. This implies that there is a pre-defined total training energy budget $E_{\text{tot}}$. It follows that a better beam alignment strategy will have a lower misalignment probability when using the same training budget $E_{\text{tot}}$.

A simple strategy of beam alignment is to train each of the $N$ beam pairs in ${\mathcal C}$ once to find the best beam pair. This is known as exhaustive search in the literature that requires $N$ measurements. Since all beam pairs look equally competitive without any prior knowledge of the channel, exhaustive search naturally allocates the same amount of energy to each measurement. Assuming that $E_{\text{tot}}$ can be divided with an arbitrarily small granularity~\footnote{This assumption holds well when the total number of pilot symbols is much larger than $N$. Considering that the bandwidth of mmWave tends to be large, even with a very short training time, the number of symbols can be very large. For instance, consider that the total training time is $10$ $\mu$s and the bandwidth is 100 MHz. This means that there are 1000 pilot symbols during the training period. Note that this $10$ $\mu$s  corresponds to roughly 13.6\% of the coherence time for a UE of relatively low-mobility at 10 m/s and the carrier frequency is 73 GHz~\cite{liu2017Jsac}.}, each measurement in exhaustive search is allocated the same amount of energy at $E = E_{\text{tot}}/N$.

\par Consider the $l$-th measurement with beam pair $({{\bf w}_l, {\bf f}_l})$ in exhaustive search. The received signal at Rx can be represented as:
\begin{align}
\mathbf{y}_l &= {\mathbf f}_l^{\dag} {\mathbf H} {\mathbf w}_l \mathbf{s} + {\mathbf f}_l^{\dag} {\mathbf Z}_l \nonumber \\
    &=  h_l \mathbf{s} + \mathbf{z}_l,~~l \in [1:N], \label{equ:received:signal}
\end{align}
where $\mathbf{s}$ is the pilot sequence of $n_s$ symbols that carries energy $E$, i.e., $\|\mathbf{s}\|_2^2 = E$, while $h_l \triangleq {\mathbf f}_l^{\dag} {\mathbf H} {\mathbf w}_l $ denotes the effective channel after Tx-Rx beamforming, ${\mathbf Z}_l \in {\mathbb C}^{N_R \times n_s}$ is the noise matrix (before Rx beamforming) with i.i.d. components $\sim {\mathcal{CN}}(0,\sigma^2)$ and thus the elements of $\mathbf{z}_l$ are i.i.d. circular Gaussian $\sim {\mathcal{CN}}(0, \sigma^2)$, given that $\left\|{\mathbf f}_l\right\|_2^2=1$.

Upon obtaining $\mathbf{y}_l$, the Rx then produces the match filtered output as:
\begin{equation}\label{equ:y:output}
r_l = \mathbf{s}^{\dag}\mathbf{y}_l = E h_l + z_l,~~l \in [1:N],
\end{equation}
where $z_l = \mathbf{s}^{\dag}\mathbf{z}_l\sim {\mathcal{CN}}(0, E\sigma^2)$. It can be seen that the SNR of the matched-filter output $r_l$ is $\frac{E |h_l|^2}{\sigma^2}=\frac{E_{\text{tot}} |h_l|^2}{N\sigma^2} $, which is proportional to the training energy allocated to each measurement.

Based on all the $N$ measurements, exhaustive search then selects the beam pair that produces the strongest match filter output:
\begin{align}
{\hat l}_{\text{ES}}= \mathop {\arg \max }\limits_{l \in [1:N]}\left |  r_l \right |. \label{equ:ES:decision}
\end{align}
If ${\hat l}_{\text{ES}}=l_{\text{opt}}$, successful alignment is attained, otherwise a misalignment is declared. It is clear that if the measurements in~\eqref{equ:y:output} were noiseless, a correct decision from~\eqref{equ:ES:decision} would always be guaranteed. However, only noisy measurements are obtained in practice and this renders exhaustive search vulnerable to potential misalignment, as is the case with any other scheme.

\par In the next section, using the same training codebook of exhaustive search, we  propose a new beam search algorithm. We shall further analyze the performance of the proposed algorithm in Section~\ref{sec:OTSS:analysis} and prove that it can outperform exhaustive search by achieving lower misalignment probability under the same training energy budget.

\section{Optimized Two-Stage Search Algorithm}\label{sec:OTSS:algorithm}

The rationale for the Optimized Two-Stage Search (OTSS) is that exhaustive search is vulnerable to false positives (i.e., beams that happen to have large match-filtered output due to noise but are not aligned with the dominant path), especially at low SNR. Using narrow beams does not necessarily help, as the narrower the beam, the less time is available for the measurement, given an overall time constraint on the beam alignment. In OTSS, we compress the exhaustive search into a shorter time period (using the same beamforming vectors), and hence use less energy for that search. We then select a set of promising beamforming directions that likely include the best direction plus the false positives, based on the initial search, and use the remaining energy to re-explore those directions. OTSS eliminates directions that have a low return in the first stage, and then focuses the residual energy in exploring the remaining directions to eliminate the false positives in the second stage. Specifically, in the second stage, an extra measurement is taken for each of the retained beam pairs and combined with the previous measurement to determine the best pair.

\par We now formally present the OTSS algorithm. Specifically, the total training energy budget $E_{\text{tot}}$ is split into two fractions $(\alpha E_{\text{tot}}, (1-\alpha)E_{\text{tot}})$, with $\alpha E_{\text{tot}}$ and $(1-\alpha)E_{\text{tot}}$ devoted to the first stage and the second stage, respectively. The splitting factor $\alpha \in (0,1]$ is to be optimized later. In the first stage, each measurement is allocated the same amount of energy:
\begin{align}
E^{(1)} = \alpha E_{\text{tot}}/N, \label{equ:1st:stage:per:beam:energy}
\end{align}
where the superscript ``(.)" indexes the stage. Considering the $l$-th measurement with beam pair $({\bf w}_l, {\bf f}_l)$ and similar to~\eqref{equ:y:output}, the matched filter output at Rx can be represented as:
\begin{align}
r_l^{(1)} = E^{(1)}h_{l} + z^{(1)}_l,~~l \in [1:N], \label{equ:y1}
\end{align}
where $h_l = {\mathbf f}_l^{\dag} {\mathbf H} {\mathbf w}_l $ as in~\eqref{equ:received:signal} and the effective noise $z^{(1)}_l\sim {\mathcal{CN}}(0, E^{(1)}\sigma^2)$ as in~\eqref{equ:y:output}. The magnitude of each matched filter output is further denoted by
\begin{align}
{\tilde T}_l^{(1)}  =  | r_l^{(1)}|,~~l \in [1:N], \label{equ:T1}
\end{align}

\par Through this exploration, the algorithm then ranks all beam pairs according to  $\{{\tilde T}_1^{(1)},\cdots, {\tilde T}_N^{(1)}\}$ in an ascending order, identifies the $K$ worst beam pairs that have the smallest value and eliminates them from further consideration. Here, $K \in [1:N-1]$ is a key algorithm parameter to be optimized. Without loss of generality, let ${\mathcal B}_{\text{G}}^{(1)}$ be the set of indices of $(N-K)$ retained beam pairs by the end of this stage.

\par In the second stage, the algorithm evenly splits the remaining training energy among beam pairs in ${\mathcal B}_{\text{G}}^{(1)}$ such that
\begin{align}
E^{(2)}= (1-\alpha) E_{\text{tot}}/(N-K), \label{equ:2nd:stage:per:beam:energy}
\end{align}
and takes an extra measurement for each pair. The matched filter output of the $l$-th beam pair in the second stage can be represented as
\begin{align}
r_l^{(2)} = E^{(2)}h_{l} + z^{(2)}_l,~~l \in {\mathcal B}_{\text{G}}^{(1)}, \label{equ:y2}
\end{align}
$z^{(2)}_l \sim {\mathcal{CN}}(0,E^{(2)}\sigma^2)$. By \textit{coherently combining} the new measurement $r_l^{(2)}$ and its previous measurement $r_l^{(1)}$, the algorithm constructs a set of combined outputs:
\begin{align}
{\tilde T}_l^{(2)} = \left | r_l^{(1)}+r_l^{(2)}\right |, ~~l \in {\mathcal B}_{\text{G}}^{(1)}. \label{equ:T2}
\end{align}
\par Finally, the beam pair with the strongest combined output is selected as the decision:
\begin{align}
{\hat l}_{\text{OTSS}}= \mathop {\arg \max }\limits_{l \in {\mathcal B}_{\text{G}}^{(1)}} {\tilde T}_l^{(2)}. \label{equ:OTSS:decision}
\end{align}

\begin{table}[t]
\begin{center} \caption{OTSS Algorithm for Beam Alignment} \label{table:algorithm}
\resizebox{0.55\textwidth}{!}{\begin{tabular}{l}
\hline
{\bf Input}: ${\mathcal C}$, beam codebook with $N$ candidate beam pairs; \\
\quad \quad \quad $E_{\text{tot}}$, total energy budget; $\alpha$, budget fraction for Stage 1;\\
\quad \quad \quad $K$, the number of beam pairs discarded in Stage 1.\\
1)~{\bf Stage 1}: \\
\quad 1.1)~{For $l \in [1:N]$}: train $({\bf w}_l, {\bf f}_l)$ to collect match-filtered measurement $r_l^{(1)}$ at Rx \\
\quad \quad~~~and compute energy ${\tilde T}_l^{(1)}$ as in~\eqref{equ:T1}.\\
\quad 1.2)~Rank all beam pairs based on their energy statistics $\{{\tilde T}_l^{(1)}\}$, \\
\quad \quad~~~discard the $K$ worst beam pairs and form survival set ${\mathcal B}_{\text{G}}^{(1)}$. \\
2)~{\bf Stage 2}: \\
\quad 2.1)~{For $l \in {\mathcal B}_{\text{G}}^{(1)}$}: retrain $({\bf w}_l, {\bf f}_l)$ to take extra measurement $r_l^{(2)}$ \\
\quad \quad~as in~\eqref{equ:y2}, and coherently combine $(r_l^{(2)},r_l^{(1)})$ to generate \\
\quad \quad~combined statistic ${\tilde T}_l^{(2)}$ as in~\eqref{equ:T2}.\\
{\bf Output}: $\hat{l}_{\text{OTSS}}= \mathop {\arg \max }\nolimits_{l \in {\mathcal B}_{\text{G}}^{(1)}} {\tilde T}_l^{(2)}$, as in~\eqref{equ:OTSS:decision}.
\\\hline
\end{tabular}}
\end{center}
\end{table}
\par The OTSS algorithm is summarized in Table~\ref{table:algorithm}. It is clear that OTSS includes exhaustive search as a special case when $\alpha =1$ and $K=N-1$. Furthermore, by properly choosing parameters $\alpha$ and $K$, it is expected that OTSS will outperform exhaustive search when the same training energy  $E_{\text{tot}}$ is used. In the next section, we will establish design guidelines on this by developing fundamental performance limits for OTSS.

\par In terms of feedback overhead, OTSS requires $\log_2{N\choose K}$ bits to select the retained beams from stage 1, and then $\log_2(N-K)$ bits to select the best beam in stage 2. As it will become clear later in the analysis, the asymptotically optimal $(N-K)$ is on the order of $\sqrt{N}$. Using  this value, we obtain an upper bound of $O(\sqrt{N} \log_2 N)$ bits in the first stage, and $O(\log_2 N)$ bits for the final stage. In contrast, exhaustive search requires $\log_2 N$ bits. Thus there is at most an increase in feedback by a factor of  $\sqrt{N}$ compared to exhaustive search. With $N=128$, exhaustive search requires $7$ bits of feedback, whereas OTSS requires $55$ bits, which is  about $8$ times more feedback than required by exhaustive search. We remark that such feedback may be obtained from a signalling channel at a lower (microwave) frequency and that it is required only for beam alignment. In return, much improved beam alignment performance can be obtained, as we demonstrate in our analytical and simulation results below. {We finally note that while OTSS requires a faster beam switching speed than exhaustive search, this faster beam switching is well within the capabilities of the state-of-the-art chip designs, e.g.,  IBM has reported beam switching speed of $<4$ ns~\cite{IBM_beamswitch}.}

\section{Performance Analysis Under Single-Path Channel Model and With Ideal Beam Codebook}\label{sec:OTSS:analysis}
\par Similar to \cite{hur2013millimeter, liu2017Jsac}, for tractability, we focus the analysis on a rank-one channel model that captures well the dominant path in a LOS environment. More general models will be numerically investigated in Section~\ref{sec:numerical:results}, and it will be shown that the insights generated from the analysis here continue to apply therein.
\par Specifically, assume both Tx and Rx adopt a uniform linear array. The rank-one channel matrix ${\mathbf H}$ is then represented as
\begin{align}
{\mathbf H} = \gamma {\mathbf u}(\phi){\mathbf v}^{\dag}(\psi),
\end{align}
where $|\gamma|^2$ is the path gain, while ${\mathbf u}(\phi)\in {\mathbb C}^{N_R \times 1}$ and ${\mathbf v}(\psi) \in {\mathbb C}^{N_T \times 1}$ are the steering vectors corresponding to AoA $\phi$ and AoD $\psi$ that are defined as
\begin{align}
&{\mathbf u}(\phi)= [1, e^{j2\pi\frac{d}{\lambda}\sin(\phi)},\cdots,e^{j2\pi\frac{d}{\lambda}(N_R-1)\sin(\phi)}]^T,\\
&{\mathbf v}(\psi)= [1, e^{j2\pi\frac{d}{\lambda}\sin(\psi)},\cdots,e^{j2\pi\frac{d}{\lambda}(N_T-1)\sin(\psi)}]^T,
\end{align}
respectively, with $\lambda$ being the wave-length and $d$ being the antenna spacing. Under this model, effective channel $h_l$ that accounts for Tx-Rx beams $({\mathbf w}_l, {\mathbf f}_l)$ as in~\eqref{equ:received:signal} is specialized to
\begin{align}
h_l  =  \gamma {\mathbf f}^{\dag}_l {\mathbf u}(\phi){\mathbf v}^{\dag}(\psi) {\mathbf w}_l,
\end{align}
and the corresponding channel gain is thus given by
\begin{align}
g_l \triangleq |h_l|^2 &=  |\gamma {\mathbf f}^{\dag}_l {\mathbf u}(\phi){\mathbf v}^{\dag}(\psi) {\mathbf w}_l|^2\\
&=|\gamma|^2 F_l(\phi) W_l(\psi), \label{equ:gain:def}
\end{align}
where we have defined $W_l(\psi) \triangleq |{\mathbf v}^{\dag}(\psi) {\mathbf w}_l|^2$ as the Tx beamforming gain at AoD $\psi$ and $F_l(\phi) \triangleq |{\mathbf f}^{\dag}_l {\mathbf u}(\phi)|^2$ as the Rx beamforming gain at AoA $\phi$.
\par In addition, as established in~\cite{liu2017Jsac,zhang2017codebook}, a desirable beam for beam training purpose should have uniform gain in its
intended coverage interval and zero leakage outside the interval. With this ideal beam assumption and supposing that all Tx (Rx) beams have equal-size non-overlapping coverage intervals that span the AoD range $\Psi$ (resp. AoA $\Phi$), the Tx (Rx) beamforming gain at $\psi \in \Psi$ (resp. $\phi \in \Phi$) is then quantified by:
\begin{align}
&W_l(\psi)  = \left\{ {\begin{array}{*{20}c}\label{Eq:ideal_pattern_Tx}
   W_T\triangleq{\frac{{4\pi }}{{\left| {\Omega_T} \right|/L_T }}}, ~~\text{if~}\psi \in \Psi_{{\bf w}_l} \\
   0,~~~~~~~~~~~~~~~~\text{otherwise}  \\
\end{array}} \right.\\
\text{and}~~&F_l(\phi)  = \left\{ {\begin{array}{*{20}c}\label{Eq:ideal_pattern_Rx}
   F_R\triangleq{\frac{{4\pi }}{{\left| {\Omega_R } \right|/L_R }}}, ~~\text{if~}\phi \in \Phi_{{\bf f}_l} \\
   0,~~~~~~~~~~~~~~~~~\text{otherwise}  \\
\end{array}} \right.,
\end{align}
where $\Psi_{{\bf w}_l}$ and $\Phi_{{\bf f}_l}$ denote the coverage interval of beam ${\bf w}_l$ and ${\bf f}_l$, while $\Omega_T$ and $\Omega_R$ are the solid angles spanned by the entire AoD range $\Psi$ and AoA range $\Phi$~\cite{constantine2005antenna}, respectively. For instance, when $\Psi= [0, 2\pi]$ and $L_T =16$ beams, we have that $\left| {\Omega_T} \right|= 4\pi$ and each beam attains constant gain $W_T= 16$ within its coverage interval.
\par With these ideal beam codebooks and under the single-path model with arbitrary AoA $\phi$ and AoD $\psi$ given, $g_l$ of~\eqref{equ:gain:def} is evaluated to
\begin{align}
g_l = \left\{ {\begin{array}{*{20}l}
   |\gamma|^2F_RW_T, ~~\text{if~}\psi \in \Psi_{{\bf w}_l}~\text{and~} \phi \in \Phi_{{\bf f}_l}.\\
   0,~~~~~~~~~~~~~\text{otherwise}.  \\
   \end{array}} \right. \label{equ:OTSS:bfgain}
\end{align}
Therefore, a perfect alignment through~\eqref{equ:perfect:alignment} simply chooses the unique beam pair with index $l_{\text opt}$ that leads to non-zero gain.
\par For the OTSS algorithm proposed, a misalignment event occurs if ${\hat l}_{\text{OTSS}} \ne l_{\text opt}$, and the probability of misalignment is thus defined as $p_{\text{miss}} = \Pr\{{\hat l}_{\text{OTSS}} \ne l_{\text opt}\}$. Without loss of optimality and for notational convenience, $l_{\text opt}=1$ is assumed. To further facilitate the analysis, we introduce the following normalized statistics that relate to $\{{\tilde T}_l^{(1)}\}$ of~\eqref{equ:T1} and $\{{\tilde T}_l^{(2)}\}$ of~\eqref{equ:T2} as
\begin{align}
&{T}_l^{(1)} \triangleq \frac{({\tilde T}_l^{(1)})^2}{\frac{\sigma^2}{2}E^{(1)}},~\forall l \in [1:N],\\
&{T}_l^{(2)} \triangleq \frac{({\tilde T}_l^{(2)})^2}{\frac{\sigma^2}{2}(E^{(1)}+E^{(2)})},~\forall l \in {\mathcal B}_{\text{G}}^{(1)}, \label{equ:def:T2}
\end{align}
and define ${T}_{(K)}^{(1)}$ as the $K$th order statistic (i.e., the $K$th smallest value) of $\{{T}_2^{(1)},\cdots, {T}_N^{(1)}\}$. By the law of total probability, $p_{\text{miss}}$ can then be expanded as
\begin{align}
p_{\text{miss}} &= \Pr\{{\hat l}_{\text{OTSS}} \ne 1\} \nonumber\\
&=\Pr\{1 \notin {\mathcal B}_{\text{G}}^{(1)} \} + \Pr\{{\hat l}_{\text{OTSS}} \ne 1~\text{and}~1 \in {\mathcal B}_{\text{G}}^{(1)}\} \\
&=\underbrace{ \Pr\{{T}_1^{(1)} < {T}_{(K)}^{(1)}\}}_{p_{\text{miss}}^{(1)}} + \underbrace{ \Pr\{ {T}_1^{(2)} < \mathop {\max }\limits_{l \in {\mathcal B}_{\text{G}}^{(1)}\backslash\{1\}} {T}_l^{(2)},~{T}_1^{(1)} \ge {T}_{(K)}^{(1)}\}}_{p_{\text{miss}}^{(2)}} \\
&= {p_{\text{miss}}^{(1)}} + {p_{\text{miss}}^{(2)}}, \label{equ:prob:miss1}
\end{align}
where ${p_{\text{miss}}^{(1)}}$ captures misalignment events that the first beam pair is eliminated in the first stage of OTSS, while ${p_{\text{miss}}^{(2)}}$ captures misalignment events that the first beam pair is not chosen at the end of the second stage, though it survives in the first stage.

\par In what follows, we proceed to study properties of relevant statistics $\{{T}_l^{(1)}, l \in [1:N]\}$ and $\{{T}_l^{(2)}, l \in {\mathcal B}_{\text{G}}^{(1)}\}$ and develop bounds on $p_{\text{miss}}$, based on which optimized $\alpha$ and $K$ are further derived.

\subsection{Bounds on the Probability of Misalignment}
\par Let $\chi_k^2 ({\lambda})$ denote a noncentral chi-squared distribution with degrees of freedom (DoFs) $k$ and noncentrality parameter $\lambda$. In the special case with $\lambda =0$, $\chi_k^2 ({0})$ becomes a central chi-squared distribution with DoFs $k$.
\begin{lemma} \label{lemma:T1}
For the OTSS proposed, under the single-path model and the ideal beam codebooks as defined in \eqref{Eq:ideal_pattern_Tx} and \eqref{Eq:ideal_pattern_Rx} , we have
\begin{align}
\left\{ {\begin{array}{*{20}l}
   {T_1^{(1)}  \sim \chi _2^2 ( {\lambda _1^{(1)} })}  \\
   {T_l^{(1)}  \sim \chi _2^2 (0),~~l \in \left[ {2:N} \right]},  \\
\end{array}} \right.
\end{align}
where ${\lambda_1^{(1)}}= \frac{2|\gamma|^2F_RW_TE^{(1)}}{\sigma^2}$, and all ${T_l^{(1)}}$'s are independent.
\end{lemma}
\begin{IEEEproof}
Following~\eqref{equ:OTSS:bfgain}, the match-filtered output $r^{(1)}_l$ as in~\eqref{equ:y1} is specialized to
\begin{align}
r_l^{(1)}  = \left\{ {\begin{array}{*{20}l}
   {\gamma \sqrt {F_R W_T}E^{(1)} + z_1^{(1)},~~~l = 1}  \\
   {z_l^{(1)}, \quad \quad \quad \quad \quad \quad \quad \quad l \in [2:N]},  \\
\end{array}} \right.
\end{align}
where each $z_l \sim {\mathcal{CN}}(0, E^{(1)}\sigma^2)$. It is immediate to show that each ${T_l^{(1)}}$ follows the distribution as given by using its definition ${T_l^{(1)}} = \frac{| {r_l^{(1)}}|^2}{\frac{\sigma^2}{2}E^{(1)}}$. In addition, ${T_l^{(1)}}$'s are mutually independent, since each of the received signals is measured at different times under different beam pairs.
\end{IEEEproof}

\par Let ${T}_{(k)}^{(1)}$ be the $k$th order statistic (i.e., the $k$th smallest value) of $\{{T}_2^{(1)},\cdots, {T}_N^{(1)}\}$. Using Lemma~\ref{lemma:T1}, it is standard to establish the result as stated in the following corollary~\cite[p.5]{balakrishnan1998handbook}.
\begin{corollary}\label{corollary:order:stat}
The probability density function of ${T}_{(k)}^{(1)}$ is
\begin{align}
f_{{T}_{(k)}^{(1)}}(x) &= \frac{{( {N - 1})!}}{2{( {k - 1})!( {N-1-k})!}}~\times \Big(1-\exp(-x/2)\Big)^{(k-1)}\exp\Big(-\frac{N-k}{2}x\Big),\nonumber\\
&~~~~~~~~~~~~~~~~x\ge 0,~k \in [1:N-1]. \label{equ:order:statistic}
\end{align}
\end{corollary}

\par With this and Lemma~\ref{lemma:T1}, we are ready to compute ${p_{\text{miss}}^{(1)}}$.
\begin{proposition} \label{prop:pmiss1}
For the OTSS proposed, misalignment probability ${p_{\text{miss}}^{(1)}} = \Pr\{{T}_1^{(1)} < {T}_{(K)}^{(1)}\}$ at Stage 1 is quantified by
\begin{align}
{p_{\text{miss}}^{(1)}} = &\frac{{( {N - 1})!}}{{( {K - 1})!( {N -1-K})!}}~\times \nonumber\\
&\sum\limits_{n = 0}^{K - 1}( { - 1})^n {\frac{{ K-1 \choose n }}{{( {N - K + n})( {N -K + n + 1})}}\exp\Big({ - \frac{{\lambda _1^{(1)} (N-K + n)}}{{2({N-K + n + 1})}}}\Big)} \label{equ:pmiss1}
\end{align}
where ${\lambda_1^{(1)}}= \frac{2|\gamma|^2F_RW_TE^{(1)}}{\sigma^2}$.
\end{proposition}
\begin{IEEEproof}
See Appendix~\ref{appendix:prop:pmiss1} for detailed proof.
\end{IEEEproof}

\begin{remark}
With the expression derived, it is trivial to see that ${p_{\text{miss}}^{(1)}}$ vanishes, when $\lambda_1^{(1)} \to \infty$ (e.g., $E_{\text{tot}} \to \infty$) on one extreme. On the other extreme with $\lambda_1^{(1)} \to 0$ (e.g., $E_{\text{tot}} \to 0$), we have that ${p_{\text{miss}}^{(1)}}$ approaches to
\begin{align}
&\frac{{( {N - 1})!}}{{( {K - 1})!( {N - K - 1})!}} \sum\limits_{n = 0}^{K - 1}( { - 1})^n \frac{{ K-1 \choose n }}{{( {N - K + n})( {N -K + n + 1})}} \\
=&\frac{{( {N - 1})!}}{{( {K - 1})!( {N - K - 1})!}} \sum\limits_{n = 0}^{K - 1}( { - 1})^n \frac{{ K-1 \choose n }\Gamma(N-K+n)}{\Gamma(N-K+2+n)} \\
=&\frac{{( {N - 1})!}}{{( {K - 1})!( {N - K - 1})!}}{\rm B}(K+1, N-K) \label{equ:pmiss1:specialcase}\\
=&\frac{{( {N - 1})!}}{{( {K - 1})!( {N - K - 1})!}} \frac{{K!({N-K-1})!}}{{{N}!}} = \frac{K}{N},
\end{align}
where~\eqref{equ:pmiss1:specialcase} follows from~\cite[Equation 0.160.2]{gradshteyn2014table} with ${\rm B}(x,y)$ and $\Gamma(z)$ being Beta and Gamma function, respectively. This result coincides with the general intuition that if no measurements were taken and $K$ out of $N$ beam pairs were randomly eliminated, the probability that the optimal beam pair is eliminated is simply $K/N$.~\remarkend
\end{remark}

\par As for Stage 2, deriving an exact characterization of ${p_{\text{miss}}^{(2)}}$ is more challenging, since energy statistic ${T}_l^{(2)}$ for each of the survived beam pairs is coupled with its previous statistic at Stage~1 through coherent energy combining~\eqref{equ:T2}.
\par In particular, for each survived $l \in {\mathcal B}_{\text{G}}^{(1)}\backslash\{1\}$, its ${T}_l^{(1)}$ must be one of $\{T_{(K+1)}^{(1)}, T_{(K+2)}^{(1)}, \cdots, T_{(N-1)}^{(1)}\}$, recalling $T_{(k)}^{(1)}$ is the $k$th order statistic of $\{{T}_2^{(1)},\cdots,{T}_{N}^{(1)}\}$. Also noting that the extra measurement under each beam-pair $l \in {\mathcal B}_{\text{G}}^{(1)}\backslash\{1\}$ at Stage 2 is simply noise, we thus define a set of auxiliary variables as
\begin{align}
T'^{(2)}_{j} = \frac{\Big|\sqrt{T_{(K+j)}^{(1)}\frac{\sigma^2}{2}E^{(1)}} + z_j^{(2)}\Big|^2}{\frac{\sigma^2}{2}{(E^{(1)}+E^{(2)})}},\label{equ:def:auxilaryRV}
\end{align}
where $z^{(2)}_j \sim {\mathcal{CN}} (0,E^{(2)}\sigma^2)$ for ${j \in [1:N-K-1]}$. It is clear that $\{T'^{(2)}_{j}, j \in [1:N-K-1]\}$ and $\{{T}_l^{(2)}, l \in {\mathcal B}_{\text{G}}^{(1)}\backslash\{1\}\}$ are statistically equivalent.
\par We now propose an upper bound on ${p_{\text{miss}}^{(2)}}$ as
\begin{align}
{p_{\text{miss}}^{(2)}} &= \Pr\{ {T}_1^{(2)} < \mathop {\max }\limits_{l \in {\mathcal B}_{\text{G}}^{(1)}\backslash\{1\}} {T}_l^{(2)},~{T}_1^{(1)} \ge {T}_{(K)}^{(1)}\} \nonumber\\
&\le \Pr\{ {T}_1^{(2)} < \mathop {\max }\limits_{l \in {\mathcal B}_{\text{G}}^{(1)}\backslash\{1\}} {T}_l^{(2)}\} \\
&= \Pr\{ {T}_1^{(2)} < \mathop {\max }\limits_{j \in [1:N-K-1]} T'^{(2)}_{j}\}  \\
&\le \sum\limits_{j=1}^{N-K-1} {\Pr \{ {T}_1^{(2)} <  T'^{(2)}_{j}\} } \label{equ:upper:union}\\
&\triangleq {\bar p_{\text{miss}}^{(2)}},
\end{align}
where~\eqref{equ:upper:union} follows from the union bound argument.

\begin{lemma}\label{lemma:Stage2}
Variable ${T}_1^{(2)} \sim \chi _2^2 ( {\lambda _1^{(2)} })$ with noncentrality parameter
\begin{align}{\lambda _1^{(2)}}= \frac{2|\gamma|^2F_RW_T(E^{(1)}+E^{(2)})}{\sigma^2},\label{equ:noncentrality}
\end{align}
while conditioned $T_{(K+j)}^{(1)} = x$, we have that
\begin{align}
\frac{E^{(1)} + E^{(2)}}{E^{(2)}}T'^{(2)}_{j} \sim \chi _2^2 \Big(\frac{E^{(1)}}{E^{(2)}}x\Big), j \in [1:N-K-1].\label{equ:conditional:chisquared}
\end{align}
\end{lemma}
\begin{IEEEproof}
The proof mainly follows by the construction of each variable and the definition of noncentral chi-squared distribution, see Appendix~\ref{appendix:lemma:Stage2}.
\end{IEEEproof}

\par With this lemma, we are ready to compute ${\bar p_{\text{miss}}^{(2)}}$.
\begin{proposition}\label{prop:pmiss2}
For the OTSS proposed, misalignment probability ${p_{\text{miss}}^{(2)}}$ at Stage 2 is upper bounded by ${\bar p_{\text{miss}}^{(2)}}$, which can be computed as
\begin{align}
{\bar p_{\text{miss}}^{(2)}}  = \sum\limits_{j = 1}^{N - K - 1} {\int_0^\infty F_{(2,2)}\Big(\frac{{E^{(2)}}}{{E^{(1)}}+{E^{(2)}}}\Big|\lambda^{(2)}_1,\frac{E^{(1)}}{E^{(2)}}x\Big)f_{T_{(K+j)}^{(1)}}(x)dx}, \label{equ:pmiss2:ub}
\end{align}
where $f_{T_{(K+j)}^{(1)}}(x)$ is given by~\eqref{equ:order:statistic} and $F_{(n_1,n_2)}(z|\eta_1,\eta_2)$ is the cumulative distribution function (CDF) of a doubly noncentral $F$-distribution $F(n_1,n_2,\eta_1,\eta_2)$ with DoFs $(n_1, n_2)$ and noncentrality parameters $(\eta_1,\eta_2)$.
\end{proposition}
\begin{IEEEproof}
It suffices to show that each ${\Pr \{ {T}_1^{(2)} <  T'^{(2)}_{j}\}}$ in ${\bar p_{\text{miss}}^{(2)}}$ can be computed as
\begin{align}
&{\Pr \{ {T}_1^{(2)} <  T'^{(2)}_{j}\}} = {\mathbb E}_{T_{(K+j)}^{(1)}}\Big[\Pr \{ {T}_1^{(2)} <  T'^{(2)}_{j}|T_{(K+j)}^{(1)}\}\Big]\nonumber\\
&={\mathbb E}_{T_{(K+j)}^{(1)}}\Big[\Pr \Big\{ \frac{{T}_1^{(2)}}{\frac{E^{(1)} + E^{(2)}}{E^{(2)}}T'^{(2)}_{j}} < \frac{E^{(2)}}{E^{(1)} + E^{(2)}} \Big|T_{(K+j)}^{(1)}\Big\}\Big] \nonumber \\
&={\mathbb E}_{T_{(K+j)}^{(1)}}\Big[F_{(2,2)}\Big(\frac{{E^{(2)}}}{{E^{(1)}}+{E^{(2)}}}\Big|\lambda^{(2)}_1,\frac{E^{(1)}}{E^{(2)}}{T_{(K+j)}^{(1)}}\Big)\Big], \label{equ:pmiss2:ub:proof}
\end{align}
where~\eqref{equ:pmiss2:ub:proof} follows from Lemma~\ref{lemma:Stage2} and the definition of doubly noncentral $F$-distribution.
\end{IEEEproof}

\par We therefore establish an upper bound on ${p_{\text{miss}}}$ of OTSS (denoted by ${\bar p_{\text{miss}}}$) as
\begin{align}
{\bar p_{\text{miss}}} = p_{\text{miss}}^{(1)} + {\bar p_{\text{miss}}^{(2)}},
\end{align}
where $p_{\text{miss}}^{(1)}$ is quantified by~\eqref{equ:pmiss1} in Proposition~\ref{prop:pmiss1}, and ${\bar p_{\text{miss}}^{(2)}}$ is quantified by~\eqref{equ:pmiss2:ub} in Proposition~\ref{prop:pmiss2}.

\par Given $N$ candidate beam pairs and the training energy budget $E_{\text{tot}}$, we can thus optimize OTSS parameters $(K, \alpha)$ such that ${\bar p_{\text{miss}}}$ established is minimized, i.e.,
\begin{align}
&\min_{K,\alpha} {\bar p_{\text{miss}}} \label{equ:upperbound:opt1}\\
&\text{subject to}~K \in [1:N-1],~\alpha \in (0,1]. \label{equ:upperbound:opt2}
\end{align}
One naive strategy of finding the optimal solution $({\bar K}^*, {\bar \alpha}^*)$ is to carry out a two-dimensional search over the feasible region, e.g., over all possible $K$'s and discretized points in $(0,1]$. However, this can be very computationally demanding especially when $N$ is large. In addition, the optimal solution is generally coupled with $N$, $E_{\text{tot}}$ and other system parameters (such as the effective channel gain), which is not a desirable feature from a practical design point of view. Given these considerations, we study the asymptotic behavior of the upper bound, with the aim to establish a simpler yet useful guideline on the choice of these two parameters of OTSS.

\subsection{Asymptotic Performance Analysis and Further Insights}
\par We focus on understanding how the upper bound decays as the training energy budget goes large, or equivalently the total number of training pilot symbols goes large.
\begin{proposition}\label{proposition:LDP}
Misalignment probability $p_{\text{miss}}^{(1)}$ satisfies a large deviation principle (LDP) with decay rate
\begin{align}
-\mathop {\lim }\limits_{E_{\text{tot}}\to \infty } \frac{1}{{E_{\text{tot}} }}\log{p_{\text{miss}}^{(1)}} = \frac{{\xi _1^{(1)} }}{{2\left( {1 + \frac{1}{N - K}} \right)}},
\end{align}
while ${\bar p_{\text{miss}}^{(2)}}$ satisfies a LDP with decay rate
\begin{align}
-\mathop {\lim}\limits_{E_{\text{tot}}\to \infty } \frac{1}{{E_{\text{tot}} }}\log{\bar p_{\text{miss}}^{(2)}}= \frac{{\xi _1^{(2)} }}{4},
\end{align}
where ${\xi _1^{(1)}} = \frac{2|\gamma|^2F_RW_T\alpha}{\sigma^2N}$ and ${\xi _1^{(2)}} = \frac{2|\gamma|^2F_RW_T}{\sigma^2}(\frac{\alpha}{N} + \frac{1-\alpha}{N-K})$. Therefore, the decay rate of ${\bar p_{\text{miss}}}$ is given by
\begin{align}
 &-\mathop {\lim }\limits_{E_{\text{tot}}\to \infty } \frac{1}{{E_{\text{tot}} }}\log {\bar p_{\text{miss}}} = \min \left\{ {\frac{{\xi _1^{(1)} }}{{2\left( {1 + \frac{1}{N - K}} \right)}},\frac{{\xi _1^{(2)} }}{4}} \right\} \triangleq I_{\bar p_{\text{miss}}}.
\end{align}
\end{proposition}
\begin{IEEEproof}
See the proof in Appendix~\ref{appendix:proposition:LDP}.
\end{IEEEproof}

\par With this result, we further derive the optimal $(K^*,\alpha^*)$ that maximize decay rate $I_{\bar p_{\text{miss}}}$.
\begin{proposition}\label{proposition:opt:decayrate}
For the OTSS proposed, the parameters $(K^*,\alpha^*)$ that maximize decay rate $I_{\bar p_{\text{miss}}}$ are given by
\begin{align}
K^* &= N - \text{round}(\sqrt{N}), \label{equ:K:star}\\
\alpha^* &= \frac{N(N-K^*+1)}{K^*(N-K^*+1)+2(N-K^*)^2}, \label{equ:alpha:star}
\end{align}
respectively, and the corresponding decay rate $I_{\bar p_{\text{miss}}}^*$ is
\begin{align}
I_{\bar p_{\text{miss}}}^*= \frac{|\gamma|^2F_RW_T}{2\sigma^2\Big(N-\frac{K^*(N-K^*-1)}{2(N-K^*)}\Big)}. \label{equ:opt:decay:rate:OTSS}
\end{align}
\end{proposition}
\begin{IEEEproof}
See Appendix~\ref{appendix:proposition:opt:decayrate}.
\end{IEEEproof}

\begin{remark} Proposition~\ref{proposition:opt:decayrate} provides a neat guideline on the choice of OTSS parameters, since the parameters $(K^*,\alpha^*)$ established here are only a function of $N$ and do not depend on the training budget and other system parameters. We shall show shortly by numerical results that OTSS with $(K^*,\alpha^*)$ performs close to OTSS with $({\bar K}^*,{\bar \alpha}^*)$ optimized under each budget $E_{\text{tot}}$ given.~\remarkend
\end{remark}

\begin{remark} Proposition~\ref{proposition:opt:decayrate} also allows us to conclude that the proposed OTSS outperforms the state-of-the-art beam search baselines under the single-path model considered.
\par Specifically, in~\cite{liu2017Jsac}, we have proved that exhaustive search asymptotically outperforms hierarchical search\footnote{Here, hierarchical search includes both the conventional case with equal training energy allocated among beam pairs examined across different stages and the optimized case with training energy allocated in a manner that equalizes the misalignment probability at different stages, see Remark~$1$ of~\cite{liu2017Jsac}.} in the sense that its misalignment probability has a larger decay rate (with respect to the training energy budget) quantified by
\begin{align}
I_{\text{ES}} = \frac{1}{N}\frac{2|\gamma|^2F_RW_T}{4\sigma^2} = \frac{|\gamma|^2F_RW_T}{2\sigma^2N}.
\end{align}
Comparing $I_{\text{ES}}$ with $I_{\bar p_{\text{miss}}}^*$ of~\eqref{equ:opt:decay:rate:OTSS}, it is immediate to see that $I_{\bar p_{\text{miss}}}^*$ is larger. Therefore, OTSS asymptotically outperforms exhaustive search and thus also hierarchical search.~\remarkend
\end{remark}

\subsection{Numerical Evaluation and Comparison}\label{sub:OTSS:comparison}
\par We now provide a numerical example to validate the insights generated from the analysis above.

\par Specifically, OTSS and exhaustive search both employ the same single-level codebook, with $L_T =16$ ideal Tx beams covering AoD range $[0,2\pi]$ and $L_R =8$ ideal Rx beams covering AoA range $[0,2\pi]$. Therefore, the beamforming gains are $W_{T}=16$ and $F_{R} = 8$. The total number of candidate beam pairs is $N=128$.
\par Hierarchical search instead employs $4$-level Tx-beam codebooks, in which $2^{k_T}$ Tx beams are used at level $k_T \in [1:4]$ to cover $[0,2\pi]$ (see, e.g., \cite{liu2017Jsac,liu2018Mag} for details on hierarchical codebook arrangement) and the last level codebook is the same as that of OTSS for fair comparison. Similarly, $3$-level Rx-beam codebooks are used with $2^{k_R}$ Rx beams at level $k_R \in [1:3]$ and the last level codebook the same as that of OTSS. Hierarchical search completes in four stages, wherein two Tx-beams and two Rx-beams are scanned at each of the first three stages, while two Tx-beams are scanned at the last stage, with the Rx-beam fixed at the best one determined from the previous stage. Therefore, Tx beamforming gain $W_T^{(k)}= 2^k $ at stage $k \in [1:4]$, while Rx beamforming gain $F_R^{(k)}= 2^k $ at stage $k \in [1:3]$ and is fixed at $8$ for the last stage. The total number of beam pairs examined by hierarchical search is $N_{\text{HS}} = 4 + 4 + 4 + 2 = 14.$

\begin{figure}[t]
\centering
\includegraphics[width=0.6\textwidth]{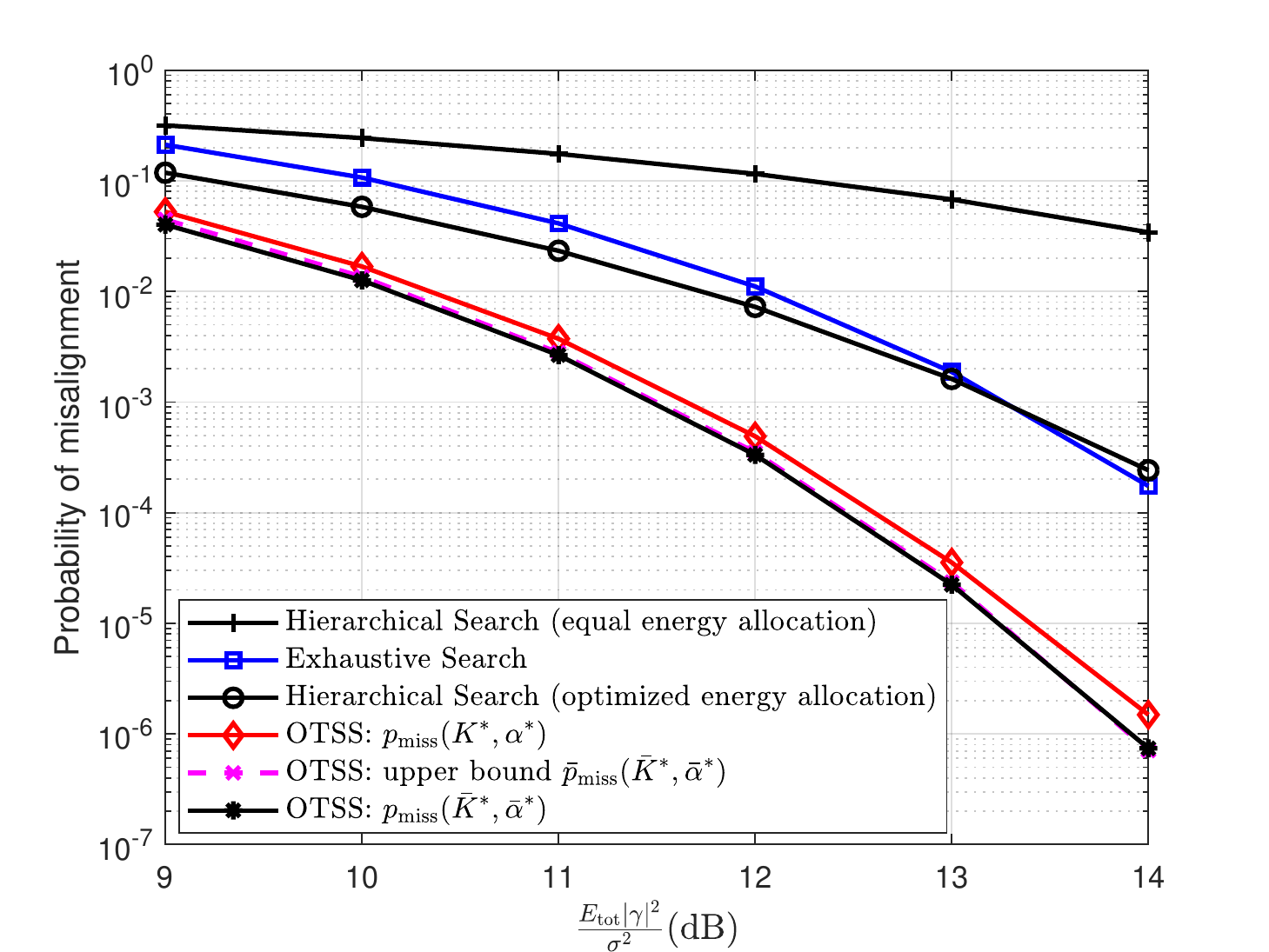}
\caption{Performance comparison of different beam alignment algorithms under single-path channel model and with ideal beam codebooks. To produce the figure, the parameters $(K^*, \alpha^*) = (117, 0.93)$, which are evaluated via~\eqref{equ:K:star} and~\eqref{equ:alpha:star} with $N=128$. The parameters $(\bar K^*, \bar \alpha^*)$ are budget-dependent and found by brute-force search to minimize ${\bar p_{\text{miss}}}$, see~\eqref{equ:upperbound:opt1}-\eqref{equ:upperbound:opt2}. In particular, $(\bar K^*, \bar \alpha^*) = \{(108,0.79), (107, 0.81), (106, 0.83), (105, 0.85), (105, 0.86), (106, 0.88)\}$ when $\frac{{E_{\rm{tot}}|\gamma|^2}}{\sigma^2} =\{9,10,11,12,13,14\}$ dB, respectively. Note that the training energy per measurement under the different schemes are different. For instance, the training energy per measurement of  exhaustive search is ${E_{\rm{tot}}}/N$, and therefore the effective signal-to-noise ratio without beamforming at Rx $\frac{{E_{\rm{tot}}}|\gamma|^2}{N\sigma^2}$ is between $[-12:-7]$ dB for the total energy budget considered. }\label{fig:perfectbeams}
\vspace{-1.5em}
\end{figure}

\par Consider a single-path channel with path gain $|\gamma|^2=1$ and vary the total training energy budget $E_{\rm{tot}}$ so that $\frac{{E_{\rm{tot}}}|\gamma|^2}{\sigma^2} \in [9:14]$ in dB unit.  Since different search strategies examine different numbers of beam pairs, the training energy per measurement in different strategies are not necessarily the same. Specifically, the training energy per measurement of  exhaustive search is ${E_{\rm{tot}}}/N$, and therefore the effective signal-to-noise ratio without beamforming at Rx $\frac{{E_{\rm{tot}}}|\gamma|^2}{N\sigma^2}$ is between $[-12:-7]$ dB for the total energy budget considered. The training energy per measurement of hierarchical search is ${E_{\rm{tot}}}/N_{\text{HS}}$, while the training energy per measurement of OTSS is determined through~\eqref{equ:1st:stage:per:beam:energy} and~\eqref{equ:2nd:stage:per:beam:energy} for Stage 1 and Stage 2, respectively, for the chosen $\alpha$ and $K$ parameters.

\par Fig.~\ref{fig:perfectbeams} compares the performance of different beam alignment algorithms in misalignment probability. In particular, when producing the curve ``OTSS: upper bound ${\bar p_{\text{miss}}}({\bar K}^*, {\bar \alpha}^*)$", for each energy budget, we search in the feasible region of $K$ and $\alpha$ and find the pair $({\bar K}^*, {\bar \alpha}^*)$ that minimizes the upper bound. These associated $({\bar K}^*, {\bar \alpha}^*)$ are then used to evaluate the misalignment probability of OTSS for the given budget (denoted by ${p_{\text{miss}}({\bar K}^*, {\bar \alpha}^*)}$). The tightness of the upper bound is clearly evident in particular when the energy budget is large. In addition, the misalignment probability of OTSS under asymptotically-optimal $(K^*,\alpha^*)$  is also evaluated (denoted by ${p_{\text{miss}}({K}^*, {\alpha}^*)}$). It can be seen that ${p_{\text{miss}}({K}^*, {\alpha}^*)}$ performs close to ${p_{\text{miss}}({\bar K}^*, {\bar \alpha}^*)}$ with only marginal loss. This confirms the utility of the asymptotic analysis.
\par Furthermore, OTSS is shown to significantly outperform all baselines considered when they use the same finite amount of training energy budget. This observation, along with the asymptotic trend we have established, confirm the superiority of OTSS.

\par Between the baselines, exhaustive search outperforms hierarchical search with equal energy allocation among beam pairs examined and also asymptotically dominates hierarchical search with optimized energy allocation across different stages, as confirmed when $\frac{{E}_{\text{tot}}|\gamma|^2}{\sigma^2}$ increases to $14$~dB in Fig.~\ref{fig:perfectbeams}.

\section{Further Numerical Studies Under Practical Considerations}\label{sec:numerical:results}
\par In this section, we further evaluate the performance of OTSS under more general channel models and with practically synthesized beams.

\subsection{Multi-Path Channel Model and Synthesized Beams}
\par Following~\cite{liu2017Jsac}, we evaluate the performance of OTSS under both line-of-sight (LOS) and non-line-of-sight (NLOS) scenarios. In particular, in the LOS scenario, a Rician channel model is adopted where the dominant component is associated with AoD $\psi$ and AoA $\phi$, both uniformly distributed in $[0, 2\pi]$, and the Rician ${\mathcal K}$-factor (i.e., the ratio of the energy of the dominant path to the sum of the energy of the scattering components) is set to $13.2$ dB~\cite{muhi2010modelling}. In the NLOS scenario, the channel is modeled as the sum of $M$ paths, each with different AoD $\psi_m$ and AoA $\phi_m$, $m=1,\cdots,M$. Each path is again assumed to be Rician, with ${\mathcal K}$-factor set to $6$ dB~\cite{samimi201628}. The number of paths is $M = \max\{1,\zeta\}$, where $\zeta$ is a Poisson random variable with mean $1.8$ and the power fractions of the $M$ paths are generated by the method in~\cite{Akdeniz2014}. In addition, in both LOS and NLOS scenarios, the average total path gain (summed over all path components) $|\bar \gamma|^2$  is assumed to be one in the simulation.

\par
\begin{figure}[t]
\centering
\includegraphics[width=0.52\textwidth]{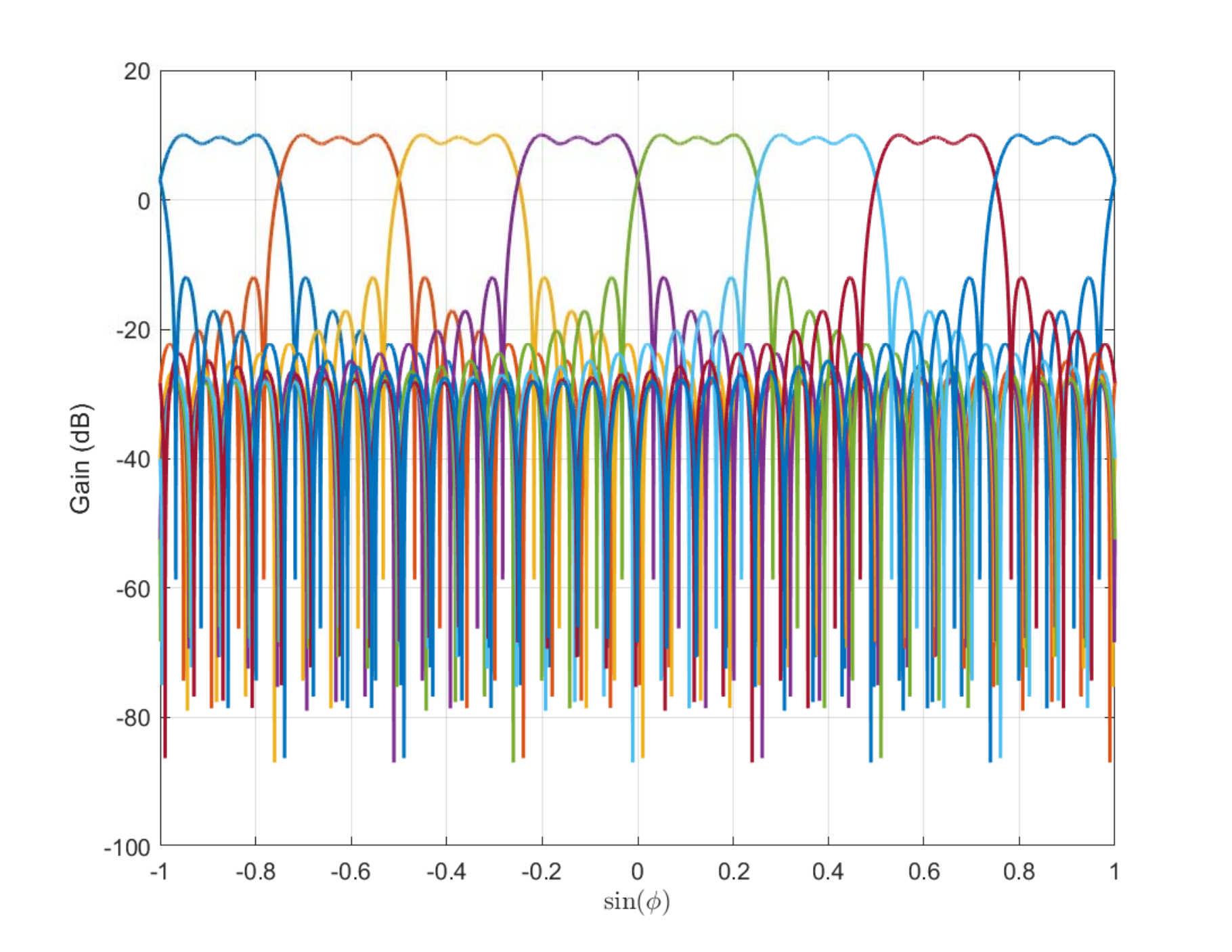}
\caption{An illustration of 8 candidate beams at Rx synthesized by the state-of-the-art flat-beam design technique~\cite{fan2018flat}.}\label{fig:practicalbeams}
\vspace{-0em}
\end{figure}

 As for beam codebooks, we consider $L_T = 16$ Tx beams and $L_R = 8$ Rx beams for exhaustive search and OTSS as in the study of Section~\ref{sub:OTSS:comparison}. These beams are now practically synthesized beams that might have attenuated gain in its passband and some leakage through its transition band and side lobe. In particular, we use the state-of-the-art flat-beam design technique~\cite{fan2018flat} to synthesize the desired Tx and Rx beams, assuming that Tx and Rx is equipped with $N_T = 64$ and $N_R =32$ antennas, respectively, and the antenna spacing is half of the wave length. Fig.~\ref{fig:practicalbeams} illustrates the Rx-beam patterns obtained for the simulation. Similarly, for hierarchical search, the same codebook setup as described in Section~\ref{sub:OTSS:comparison} is used but with synthesized flat beams for each search level.

\subsection{Simulation Results and Discussions}

\par
\begin{figure}
\centering
\begin{minipage}{.48\linewidth}
\centering
\includegraphics[width=1\textwidth]{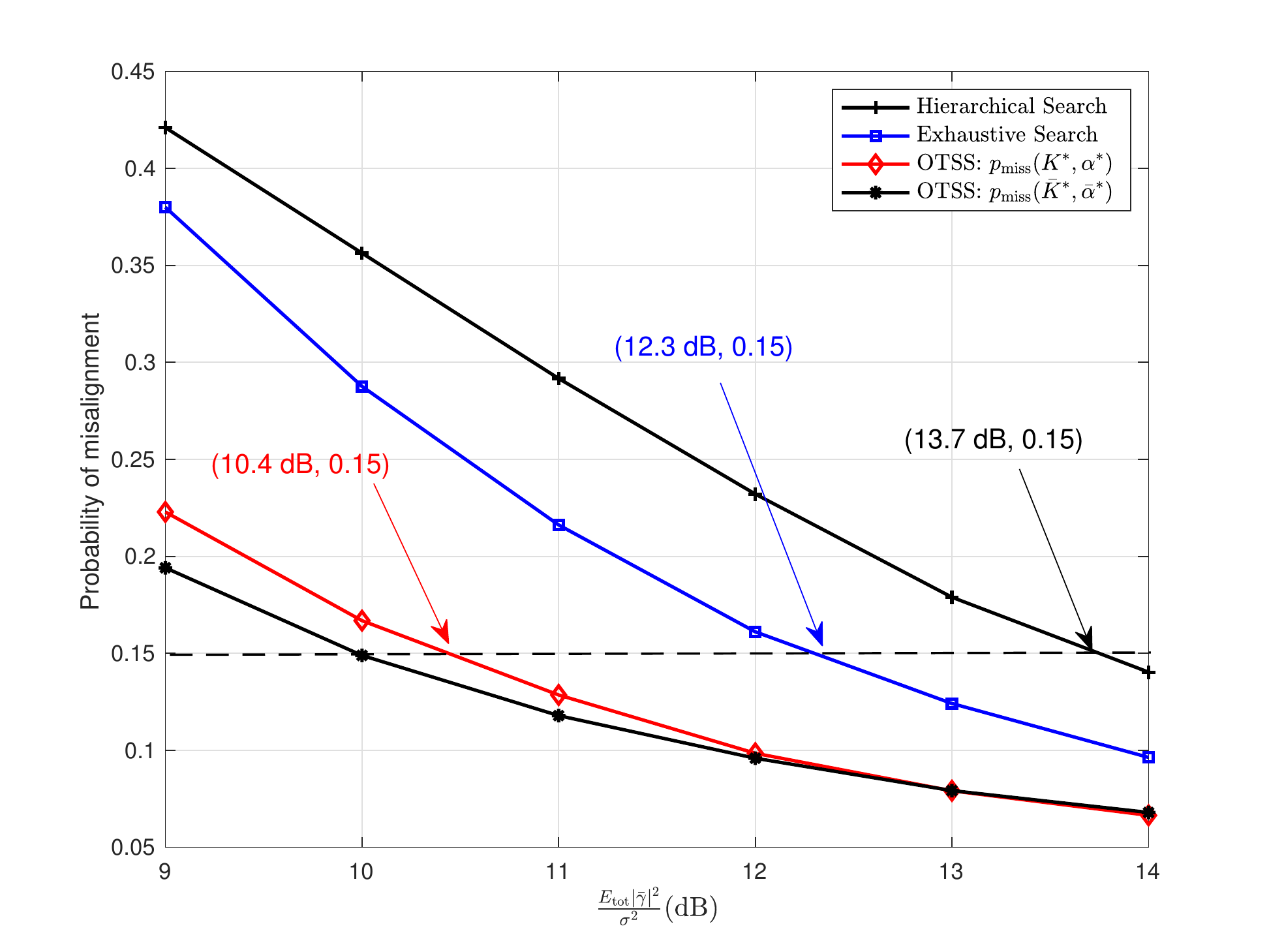}
\caption{Misalignment probability of different schemes under the LOS channel model and with practical imperfect beam codebooks. Parameters $(K^*, \alpha^*)$ and budget-dependent parameters $({\bar K}^*, {\bar \alpha}^*)$ are the same to the ones in Fig.~\ref{fig:perfectbeams}.}\label{fig:pmiss:practicalbeams}
\end{minipage}
\hspace{.05\linewidth}
\begin{minipage}{.45\linewidth}
\centering
\vspace{-0.5em}
\includegraphics[width=1.0\textwidth]{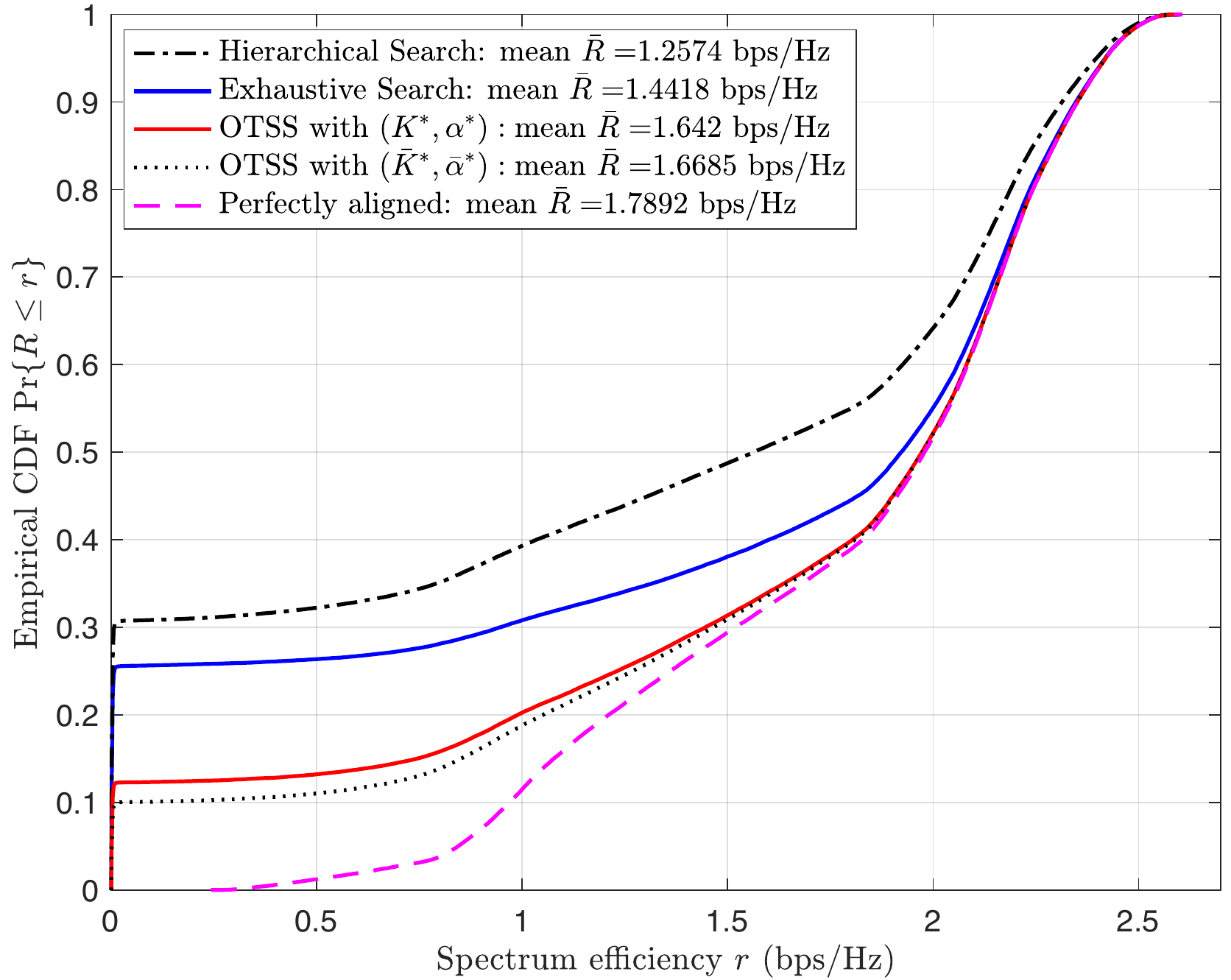}
\caption{Achievable spectrum efficiency of different schemes under the LOS channel model and with practical imperfect beam codebooks.}\label{fig:SE:practicalbeams}
\end{minipage}
\vspace{-1.5em}
\end{figure}

 We first evaluate the performance of OTSS under the LOS channel model and practical beam codebooks described. We also compare its performance with that of exhaustive search and hierarchical search (with equal energy allocation). Fig.~\ref{fig:pmiss:practicalbeams} plots the misalignment probability of the different schemes averaged through $10^5$ channel realizations. It can be seen that the relative performance trend observed under rank-one channel and ideal beams, as presented in Fig.~\ref{fig:perfectbeams}, still remains in this more practical setup:~OTSS with $(K^*, \alpha^*)$ outperforms hierarchical search and exhaustive search, while OTSS with $({\bar K}^*, {\bar \alpha}^*)$ leads to additional improvement in the regime of small to moderate training energy budget.

\par {The results in Fig.~\ref{fig:pmiss:practicalbeams} also show that OTSS can achieve the same misalignment probability target faster. As marked in Fig.~\ref{fig:pmiss:practicalbeams},  OTSS requires $\frac{E_{\text{tot}}|\bar{\gamma}|^2}{\sigma^2} = \frac{N_{\text{tot}}P_T|\bar{\gamma}|^2}{\sigma^2} = 10.4$ dB to reach a misalignment probability target of 0.15. This quantity is 12.3 dB for exhaustive search and 13.7 dB for hierarchical search. This means that to reach the same misalignment performance of 0.15, OTSS requires $N_{\text{tot}}$ which is 10.4-12.3 dB = -1.9 dB = 64.57\% of that required by exhaustive search, and 10.4-13.7 = -3.3 dB = 46.77\% of that required by hierarchical search. In other words, OTSS only requires 64.57\% of the time required by exhaustive search and 46.77\% of the time required by hierarchical search to reach the same performance. }

\par We next evaluate the spectrum efficiency for subsequent data transmission using the final Tx-Rx beam pair determined under different alignment schemes when $\frac{E_{\text{tot}}|\bar \gamma|^2}{\sigma^2}=10$ dB. Specifically, under any channel realization ${\bf H}$ generated, let $({\bf w}_{\hat l}, {\bf f}_{\hat l})$ be the TX-RX beam pair selected by a scheme under consideration. Then the corresponding spectrum efficiency $R({\bf H})$ is evaluated via Shannon's formula:
\begin{align}
R({\bf H}) = \log_2\left(1 + \frac{P_t}{\sigma^2}|{\bf f}_{\hat l}^{\dag} {\bf H}{\bf w}_{\hat l}|^2\right),
\end{align}
where $P_t$ is the transmit power. The spectrum efficiency is further compared with that of the ``perfectly aligned" benchmark, in which the optimal beam pair $({\bf w}_{l_{\text{opt}}}, {\bf f}_{l_\text{opt}})$ that maximizes the effective channel gain after beamforming is used (see~\eqref{equ:perfect:alignment}).

\par Fig.~\ref{fig:SE:practicalbeams} plots the empirical cumulative distribution function (CDF) of $R$, i.e., $\Pr\{R\le r\}$, with an averaged pre-beamforming SNR at the Rx of ${P_t}|\bar{\gamma}|^2/{\sigma^2} = -16$ dB under different schemes. Consistent with the misalignment probability comparison, OTSS also outperforms exhaustive search and hierarchial search with respect to achievable spectrum efficiency. In particular, in terms of average spectrum efficiency ($\bar R \triangleq {\mathbb E}[R]$), exhaustive search achieves $1.4418$ bps/Hz, while OTSS with $(K^*, \alpha^*)$ achieves $1.642$ bps/Hz, leading to about $(1.642-1.4418)/1.4418 = 13.9\%$ improvement over exhaustive search. In addition, considering $r = 0.5$ bps/Hz as the outage threshold, exhaustive search leads to $26.4\%$ outage probability, while OTSS only leads to $13.2\%$ outage probability.

\begin{figure}
\centering
\begin{minipage}{.45\linewidth}
\centering
\includegraphics[width=1.05\textwidth]{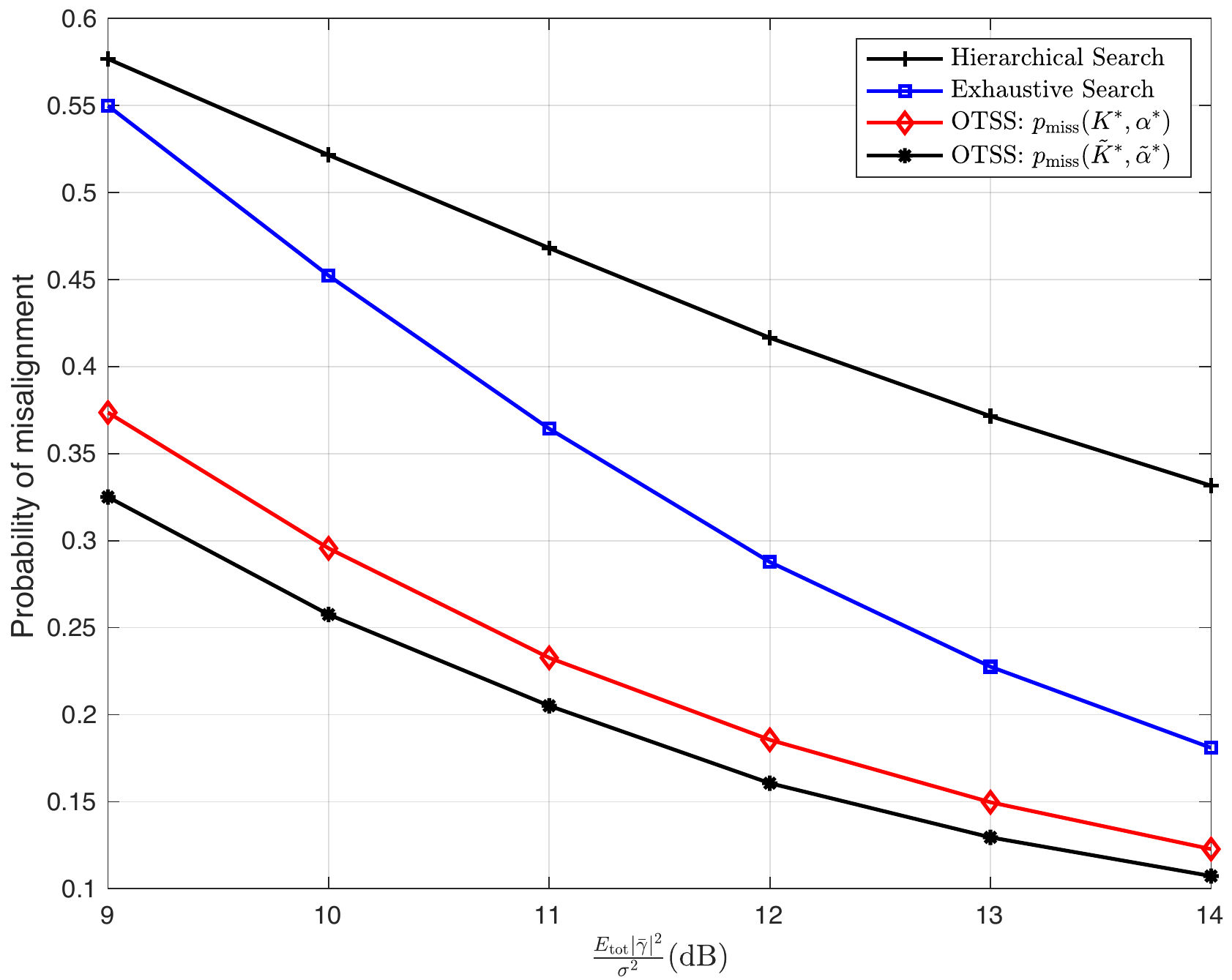}
\caption{Misalignment probability of different schemes under the NLOS channel model and with practical imperfect beam codebooks.}\label{fig:pmiss:practicalbeams:NLOS}
\end{minipage}
\hspace{.05\linewidth}
\begin{minipage}{.45\linewidth}
\centering
\includegraphics[width=\textwidth]{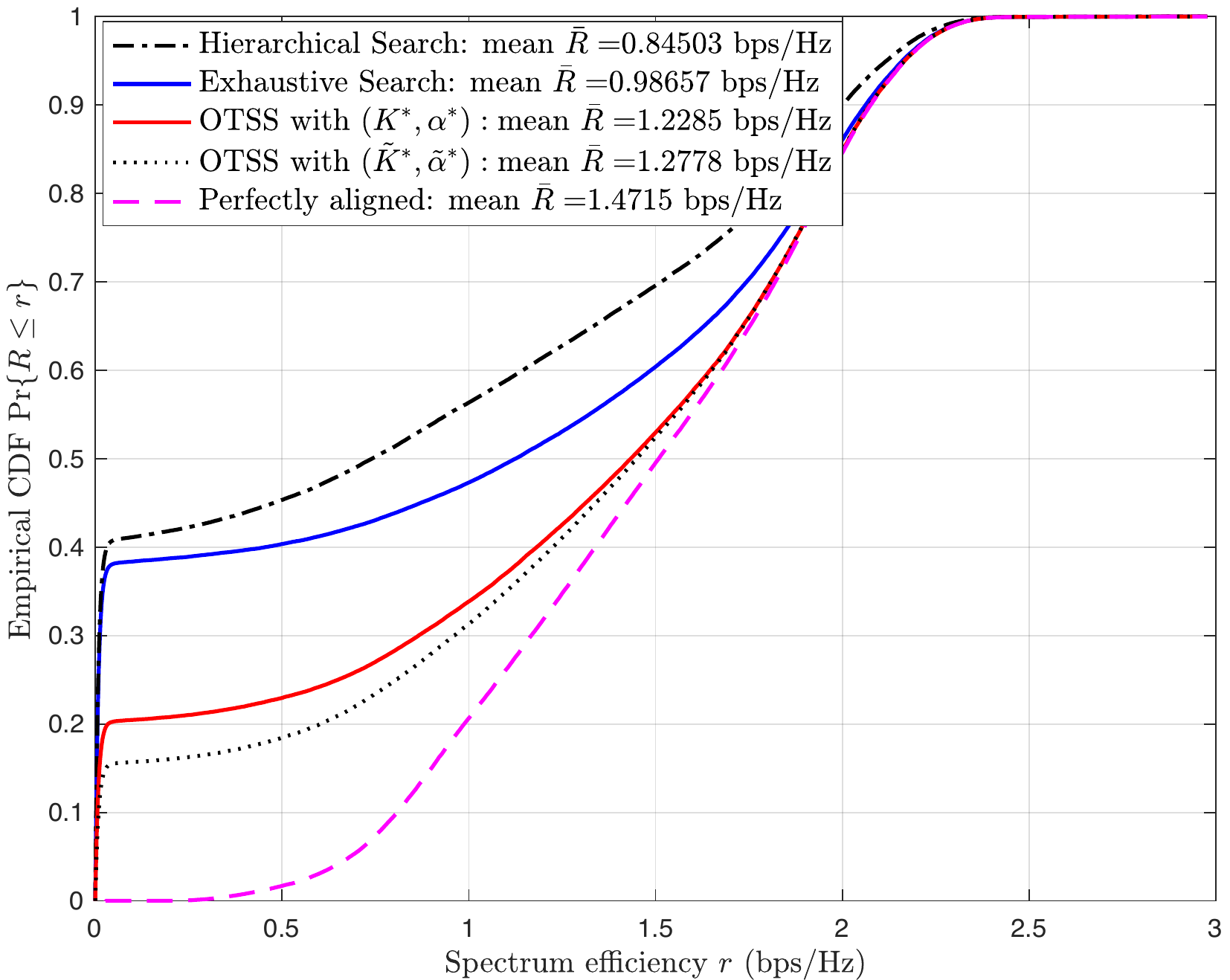}
\caption{Achievable spectrum efficiency of different schemes under the NLOS channel model and with practical imperfect beam codebooks.}\label{fig:SE:practicalbeams:NLOS}
\end{minipage}
\vspace{-1.5em}
\end{figure}

\par We next compare the performance of different schemes under the NLOS channel model. When evaluating OTSS, we also consider two design choices for the parameters $(K, \alpha)$: \textit{(i)} asymptotically optimal $(K^*,\alpha^*)$ that was derived under the single-path model; \textit{(ii)} budget-dependent $({\tilde K}^*,{\tilde \alpha}^*)$ that was found by brute-force search to minimize the average misalignment probability for a given energy budget by numerical simulations. In particular, $({\tilde K}^*,{\tilde \alpha}^*) = \{(106,0.72), (107, 0.73), (111, 0.82), (108, 0.72), (114, 0.80), (108, 0.71)\}$ when $\frac{{E_{\rm{tot}}|{\bar \gamma}|^2}}{\sigma^2} =\{9,10,11,12,13,14\}$ dB, respectively. Fig.~\ref{fig:pmiss:practicalbeams:NLOS} compares the misalignment probability of the different schemes with $\frac{E_{\text{tot}}|\bar \gamma|^2}{\sigma^2} \in [9:14]$ dB, while Fig.~\ref{fig:SE:practicalbeams:NLOS} depicts the CDF of the corresponding achievable spectrum efficiency for data transmission by using the final Tx-Rx beam pair determined under different schemes, with training energy budget fixed at $\frac{E_{\text{tot}}|\bar \gamma|^2}{\sigma^2}=10$ dB for beam alignment. It can be seen that the relative performance behavior of the schemes under the NLOS scenario is similar to that under the LOS scenario. In particular, OTSS outperforms all baselines in both misalignment probability and achievable spectrum efficiency. The parameters $(K^*,\alpha^*)$, though being derived under the single-path model, remain as remarkably excellent choices for OTSS under the realistic NLOS scenario simulated.

\par It is finally remarked that all beam-alignment schemes considered here suffer performance loss, as compared to the case with perfect beam pattern in Section~\ref{sub:OTSS:comparison}. One main source of the performance degradation is due to the overlapped transition bands of two adjacent beams (see Fig.~\ref{fig:practicalbeams}). When either AoA or AoD realization falls into such an overlapped interval, it becomes extremely difficult to make a correct decision based on noisy training measurements of two comparably strong beams. In addition, the side lobes of these imperfect beams introduce additional randomness into beam training measurements and affect the final performance. Given any finite number of Tx or Rx antennas, it is impossible to synthesize perfect beams as desired and hence these negative impacts are inevitable. Some potential remedies might include beam training with dynamic beam codebooks, e.g., intelligently shifting beams or combining adjacent beams as the search progresses, to ensure AoA or AoD realizations are always covered by the main lobe of some beam. However, such designs are beyond the scope of the current work.

\section{Conclusions}
\par In this paper, we have proposed a new Optimized Two-Stage Search (OTSS) algorithm for mmWave beam alignment. In contrast to exhaustive search where the training energy budget is equally allocated among all candidate beam pairs, the proposed OTSS first explores all beam pairs with a fraction of the budget, identifies a subset of potentially good beam directions and then spends the remaining budget on these directions to determine the best one. Fundamental analysis has been developed to establish the optimal choice for the design parameters of OTSS under a single-path channel model with ideal beam codebooks. It has been further proved that OTSS outperforms state-of-the-art algorithms (including hierarchical search and exhaustive search) when the training energy budget goes large. {OTSS has been numerically shown to achieve better performance than hierarchical search and exhaustive search when the same limited training energy/time is used under a more general channel model and practical beam codebooks. In other words, OTSS achieves the same performance faster. } As for future work, it is of great interest to establish a new analytical model that captures the essence of imperfect beam patterns and to develop new alignment algorithms with improved robustness to more practical channel models and imperfect beam scenarios. In addition, for a fixed Tx/Rx antenna geometry, it is also of importance to establish some design guidelines on the optimal choice of Tx/Rx beamwidth used for beam alignment under a given system design target, e.g., maximizing the average achievable spectrum efficiency, minimizing the outage probability, or achieving some tradeoff between these performance metrics.
\appendices
\section{Proof of Proposition~\ref{prop:pmiss1}} \label{appendix:prop:pmiss1}
\par Misalignment probability ${p_{\text{miss}}^{(1)}}$ can be represented as
\begin{align}
{p_{\text{miss}}^{(1)}} &= \Pr\{{T}_1^{(1)} < {T}_{(K)}^{(1)}\}  \\
&= \int_0^\infty \Big[1 - Q_1\Big(\sqrt{\lambda^{(1)}_1}, \sqrt{x}\Big)\Big] f_{T_{(K)}^{(1)}}(x)dx \label{equ:pmiss1:b1} \\
&= \frac{{( {N - 1})!}}{{2( {K - 1})!( {N -1-K})!}} \times \nonumber\\
&~~\int_0^\infty \Big[1-\exp{\left(-\frac{x}{2}\right)}\Big]^{K-1}\exp{\left(-\frac{N-K}{2}x\right)}\Big[1 - Q_1\Big(\sqrt{\lambda^{(1)}_1}, \sqrt{x}\Big)\Big] dx \label{equ:pmiss1:b2}\\
&=\frac{{( {N - 1})!}}{{2( {K - 1})!( {N -1-K})!}}~\times \nonumber\\
& \Bigg[ \underbrace{\int_0^\infty \Big[1-\exp{\left(-\frac{x}{2}\right)}\Big]^{K-1} \exp{\left(-\frac{N-K}{2}x\right)} dx}_{(a)}  \nonumber\\
&\hspace{-2em}-\underbrace{\int_0^\infty \Big[1-\exp{\left(-\frac{x}{2}\right)}\Big]^{K-1} \exp{\left(-\frac{N-K}{2}x\right)} Q_1\Big(\sqrt{\lambda^{(1)}_1}, \sqrt{x}\Big)dx}_{(b)}\Bigg], \label{equ:pmiss1:b2a}
\end{align}
where~\eqref{equ:pmiss1:b1} follows from Lemma~\ref{lemma:T1} and the cumulative density function of a noncentral chi-squared distribution (as a function of the Marcum $Q$-function) and~\eqref{equ:pmiss1:b2} follows from Corollary~\ref{corollary:order:stat}.
\par Now note that
\begin{align}
(a) &= {\int_0^\infty \Big[1-\exp{\left(-\frac{x}{2}\right)}\Big]^{K-1} \exp{\left(-\frac{N-K}{2}x\right)}dx}\nonumber\\
&= \sum\limits_{n = 0}^{K - 1} {( - 1)^n {{K-1}\choose{n}} \int_0^\infty \exp{\left(-\frac{N-K+n}{2}x\right)}dx} \\
&= \sum\limits_{n = 0}^{K - 1} ( - 1)^n {{K-1}\choose{n}}\frac{2}{N-K+n},
\end{align}
while
\begin{align}
&\hspace{-1em}(b)={\int_0^\infty \Big[1-\exp{\left(-\frac{x}{2}\right)}\Big]^{K-1} \exp{\left(-\frac{N-K}{2}x\right)} Q_1\Big(\sqrt{\lambda^{(1)}_1}, \sqrt{x}\Big)dx}\nonumber\\
\hspace{-1em}=&\sum\limits_{n = 0}^{K - 1} {( - 1)^n {{K-1}\choose{n}} \int_0^\infty \exp{\left(-\frac{N-K+n}{2}x\right)}Q_1\Big(\sqrt{\lambda^{(1)}_1}, \sqrt{x}\Big)dx} \nonumber \\
\hspace{-1em}=&\sum\limits_{n = 0}^{K - 1} {( - 1)^n {{K-1}\choose{n}}\Bigg[\frac{2}{N-K+n}}-\frac{2}{(N-K+n)(N-K+n+1)}\exp\Big(\frac{-\lambda_{1}^{(1)}(N-K+n)}{2(N-K+n+1)}\Big)\Bigg], \label{equ:pmiss1:b3}
\end{align}
where~\eqref{equ:pmiss1:b3} uses the fact~\cite[Equation (16)]{cui2012two} that
\begin{align}
\int_0^\infty \exp{(-px)}Q_{m}(a,b\sqrt{x})dx =\frac{1}{p}-\frac{1}{p}\left(\frac{b^2}{b^2+2p}\right)^m \exp{\left(-\frac{a^2p}{b^2+2p}\right)},
\end{align}
where $Q_{m}(a,b\sqrt{x})$ is the generalized Marcum Q-function. Plugging $(a)$ and $(b)$ above into~\eqref{equ:pmiss1:b2a} thus establishes a characterization of ${p_{\text{miss}}^{(1)}}$ as given by~\eqref{equ:pmiss1}.

\section{Proof of Lemma~\ref{lemma:Stage2}} \label{appendix:lemma:Stage2}
\par Under the single-path channel model and perfect beam codebooks considered, ${T}_1^{(2)}$ as in~\eqref{equ:def:T2} is specialized to
\begin{align}
&{T}_1^{(2)} = \frac{\left |\gamma \sqrt {F_R W_T} (E^{(1)} + E^{(2)}) + z_1^{(1)}+ z_1^{(2)} \right|^2}{\frac{\sigma^2}{2}(E^{(1)}+E^{(2)})} \nonumber\\
&=\Big |\gamma \sqrt{\frac{2F_R W_T(E^{(1)} + E^{(2)})}{\sigma^2}} + \underbrace{\frac{{z}_1^{(1)}+ {z}_1^{(2)}}{\sqrt{\frac{\sigma^2}{2}(E^{(1)}+E^{(2)})}}}_{z} \Big|^2,\nonumber
\end{align}
where $z \sim {\mathcal{CN}}(0,2)$, since $z_1^{(1)}$ $\sim {\mathcal{CN}}(0,E^{(1)}\sigma^2)$ and $z_1^{(2)}$ $\sim {\mathcal{CN}}(0, E^{(2)}\sigma^2)$ and they are independent. It is immediate to conclude that ${T}_1^{(2)}$ follows a noncentral chi-squared distribution with 2 DoFs and noncentrality parameter ${\lambda _1^{(2)}}$ as given in~\eqref{equ:noncentrality}.
\par As for $T'^{(2)}_{j}$ in~\eqref{equ:def:auxilaryRV}, conditioned on $T_{(K+j)}^{(1)} = x$,
\begin{align}
\frac{E^{(1)} + E^{(2)}}{E^{(2)}}T'^{(2)}_{j} &= \frac{\Big|\sqrt{x\frac{\sigma^2}{2}E^{(1)}} + z^{(2)}_j\Big|^2}{\frac{\sigma^2}{2}{E^{(2)}}}=\Big|\sqrt{\frac{E^{(1)}}{E^{(2)}}x} + \underbrace{\frac{z^{(2)}_j}{\sqrt{{\frac{\sigma^2}{2}{E^{(2)}}}}}}_{z'}\Big|^2,
\end{align}
where $z' \sim {\mathcal{CN}}(0,2)$, since $z^{(2)}_j \sim {\mathcal{CN}}(0,E^{(2)}\sigma^2)$. It is immediate to conclude that $\frac{E^{(1)} + E^{(2)}}{E^{(2)}}T'^{(2)}_{j}$ follows a noncentral chi-squared distribution with 2 DoFs and noncentrality parameter $\frac{E^{(1)}}{E^{(2)}}x$, given that $T_{(K+j)}^{(1)} = x$.

\section{Proof of Proposition~\ref{proposition:LDP}} \label{appendix:proposition:LDP}

\par {(\textbf{Part 1})}: We first consider $p_{\text{miss}}^{(1)} = \Pr\{{T}_1^{(1)} < {T}_{(K)}^{(1)}\}$, where we recall that ${T}_1^{(1)} \sim \chi _2^2 ( {\lambda _1^{(1)} })$ and ${T}_{(K)}^{(1)}$ is the $K$th order statistic of $(N-1)$ i.i.d. variables$\sim \chi _2^2 (0)$ as shown by Lemma~\ref{lemma:T1}.

\par Define ${\bar T}_1^{(1)} \triangleq \frac{{T}_1^{(1)}}{E_{\text{tot}}}$ and ${\bar T}_{(K)}^{(1)} \triangleq \frac{{T}_{(K)}^{(1)}}{E_{\text{tot}}}$. We thus have $p_{\text{miss}}^{(1)} = \Pr\{{T}_1^{(1)} < {T}_{(K)}^{(1)}\} = \Pr\{{\bar T}_1^{(1)} < {\bar T}_{(K)}^{(1)}\}$. Since ${\bar T}_1^{(1)}$ and ${\bar T}_{(K)}^{(1)}$ are independent, we can proceed by deriving the decay rate function for each and then combining them to characterize the decay rate of $p_{\text{miss}}^{(1)}$.

\par Specifically, we first check that the normalized logarithmic Moment Generating Function (MGF) of ${\bar T}_1^{(1)}$ exists as a extended real number as
\begin{align}
\Lambda_{{{\bar T}_1^{(1)}}}(t_1)=&\mathop {\lim }\limits_{E_{\text{tot}} \to \infty} \frac{1}{{E}_{\text{tot}}}\log{\mathbb E}[ {e^{E_{\text{tot}}t_1 {\bar T}_1^{(1)}}}] \\
=&\mathop {\lim }\limits_{E_{\text{tot}} \to \infty} \frac{1}{{E}_{\text{tot}}}\log{{\mathbb E}[ {e^{{T}_1^{(1)}t_1}}]}\\
=&\left\{{\begin{array}{*{20}c}
   \mathop {\lim }\limits_{E_{\text{tot}} \to \infty} \frac{1}{{E}_{\text{tot}}}\log\bigg(\frac{\exp\Big({\frac{\lambda^{(1)}_1 t_1}{1-2t_1}}\Big)}{1-2t_1}\bigg) = \frac{{\xi_{1}^{(1)} t_1 }}{{1 - 2t_1 }},\text{~if~}t_1 < \frac{1}{2}\\
   +\infty, \quad \quad \quad \quad \quad \quad \quad \quad \quad \quad \quad \quad \quad \quad \text{otherwise},
\end{array}} \right.  \label{equ:LDP:T1}
\end{align}
where~\eqref{equ:LDP:T1} follows from ${T}_1^{(1)} \sim \chi _2^2 ( {\lambda _1^{(1)} })$, and we have introduced $\xi_{1}^{(1)}\triangleq \frac{\lambda_1^{(1)}}{E_{\text{tot}}}= \frac{2|\gamma|^2F_RW_T\alpha}{\sigma^2N}$. Further, the origin belongs to the interior of ${\mathcal D}_{\Lambda}= \{t_1: \Lambda_{{{\bar T}_1^{(1)}}}(t_1) < \infty\}$. Therefore, ${\bar T}_1^{(1)}$ satisfies the Gartner-Ellis conditions~\cite[Assumption 2.3.2]{dembo2009large}. The rate function of ${\bar T}_1^{(1)}$ can be calculated as the Fenchel-Legendre transform of $\Lambda_{{{\bar T}_1^{(1)}}}(t_1)$~\cite[Section~2.3]{dembo2009large}:
\begin{align}
I_{{\bar T}_1^{(1)}}(u) &= \mathop {\sup}\limits_{t_1  \in {\mathbb R}} \{ ut_1  - \Lambda(t_1)\} \\
&= \mathop {\sup}\limits_{t_1 < \frac{1}{2}} \Big\{ ut_1  - \frac{{\xi_{1}^{(1)} t_1 }}{{1 - 2t_1 }}\Big\}\\
&= \frac{\Big(\sqrt{u}- \sqrt{\xi_{1}^{(1)}}\Big)^2}{2},~~u\ge 0.
\end{align}
\par Similarly, for ${\bar T}_{(K)}^{(1)}$, its normalized logarithmic MGF exists as an extended real number as
\begin{align}
\Lambda_{{{\bar T}_{(K)}^{(1)}}}(t_2) =&\mathop {\lim }\limits_{E_{\text{tot}} \to \infty} \frac{1}{{E}_{\text{tot}}}\log{\mathbb E}[ {e^{E_{\text{tot}}t_2 {\bar T}_{(K)}^{(1)}}}] \\
=&\mathop {\lim }\limits_{E_{\text{tot}} \to \infty} \frac{1}{{E}_{\text{tot}}}\log{{\mathbb E}[ {e^{{T}_{(K)}^{(1)}t_2}}]}\\
=&\mathop {\lim }\limits_{E_{\text{tot}} \to \infty} \frac{1}{{E}_{\text{tot}}}\log{{\mathbb E}[ {e^{\sum\nolimits_{j = 1}^K {Z_j}t_2}}]} \label{equ:Korder:stat:exp}\\
=&\left\{{\begin{array}{*{20}c}
   \mathop {\lim }\limits_{E_{\text{tot}} \to \infty} \frac{1}{{E}_{\text{tot}}}\log\Big(\prod\limits_{j = 1}^K {\frac{{\frac{N - j}{2}}}{{\frac{N - j}{2}- t_2 }}}\Big) = 0,\text{~if~}t_2 < \frac{N-K}{2}\\
   +\infty, \quad \quad \quad \quad \quad \quad \quad \quad \quad \quad \quad \quad \text{otherwise},
\end{array}} \right.  \label{equ:LDP:T2}
\end{align}
where~\eqref{equ:Korder:stat:exp} uses the fact that ${T}_{(K)}^{(1)}$ (the $K$th order statistic of $(N-1)$ i.i.d. variables$\sim \chi _2^2 (0)$) is statistically equivalent to $\sum\nolimits_{j = 1}^K {Z_j}$, where $Z_j$ is distributed as an exponential distribution with rate parameter $\frac{N-j}{2}$ (i.e., $Z_j \sim \text{Exp}(\frac{N-j}{2})$)~\cite[Chapter 1]{balakrishnan1998handbook}, while~\eqref{equ:LDP:T2} follows from the MGF of an exponential distributed variable. Further, the origin belongs to the interior of ${\mathcal D}_{\Lambda}= \{t_2: \Lambda_{{{\bar T}_{(K)}^{(1)}}}(t_2) < \infty\}$. Therefore, ${\bar T}_{(K)}^{(1)}$ satisfies the Gartner-Ellis conditions~\cite[Assumption 2.3.2]{dembo2009large}. The rate function of ${\bar T}_{(K)}^{(1)}$ can be calculated as the Fenchel-Legendre transform of $\Lambda_{{{\bar T}_{(K)}^{(1)}}}(t_2)$:
\begin{align}
I_{{\bar T}_{(K)}^{(1)}}(v) &= \mathop {\sup}\limits_{t_2  \in {\mathbb R}} \{ vt_2  - \Lambda_{{{\bar T}_{(K)}^{(1)}}}(t_2)\} \\
&= \mathop {\sup}\limits_{t_2 < \frac{N-K}{2}} \{ vt_2\}\\
&= \frac{N-K}{2}v,~~v\ge 0.
\end{align}
\par Since both $\Lambda_{{\bar T}_1^{(1)}}(t_1)$ and $\Lambda_{{{\bar T}_{(K)}^{(1)}}}(t_2)$ are essentially smooth~\cite[page~44]{dembo2009large}, lower semicontinuous functions, the large deviation principle (LDP) holds for both ${\bar T}_1^{(1)}$ and ${{\bar T}_{(K)}^{(1)}}$ with the good rate function $I_{{\bar T}_1^{(1)}}(u)$ and $I_{{\bar T}_{(K)}^{(1)}}(v)$, respectively~\cite[Theorem 2.3.6 (Gartner-Ellis)]{dembo2009large}. We thus further have
\begin{align}
&-\mathop {\lim }\limits_{E_{\text{tot}}\to \infty }\frac{1}{{E_{\text{tot}} }} \log {p_{\text{miss}}^{(1)}} \nonumber\\
=&-\mathop {\lim }\limits_{E_{\text{tot}}\to \infty }\frac{1}{{E_{\text{tot}} }} \log {\Pr\{{\bar T}_1^{(1)} < {\bar T}_{(K)}^{(1)}\}}\\
=&\mathop {\inf}\limits_{0\le u \le v} \{I_{{\bar T}_1^{(1)}}(u) + I_{{\bar T}_{(K)}^{(1)}}(v)\} \\
=&\mathop {\inf}\limits_{0\le u \le v} \Bigg\{\frac{\Big(\sqrt{u}- \sqrt{\xi_{1}^{(1)}}\Big)^2}{2} + \frac{N-K}{2}v\Bigg\}. \label{equ:minimization:p1}
\end{align}
\par Using the Karush-Kuhn-Tucker conditions~\cite{boyd2004convex} for the minimization problem in~\eqref{equ:minimization:p1}, it can be shown that the infimum is attained when $u^*= v^* = \frac{\xi_{1}^{(1)}}{(N-K+1)^2}$. Therefore, the decay rate of ${p_{\text{miss}}^{(1)}}$ is given by
\begin{align}
-\mathop {\lim }\limits_{E_{\text{tot}}\to \infty }\frac{1}{{E_{\text{tot}} }} \log {p_{\text{miss}}^{(1)}}&= \frac{\Big(\sqrt{u^*}- \sqrt{\xi_{1}^{(1)}}\Big)^2}{2} + \frac{N-K}{2}v^* \nonumber\\
&=\frac{\xi_{1}^{(1)}}{2\big(1+\frac{1}{N-K}\big)}.
\end{align}

\par {(\textbf{Part 2})}: Now consider ${\bar p_{\text{miss}}^{(2)}} = \sum\limits_{j=1}^{N-K-1} {\Pr \{ {T}_1^{(2)} <  T'^{(2)}_{j}\} }$, where recall that ${T}_1^{(2)} \sim \chi _2^2 ( {\lambda _1^{(2)} })$ and a statistical property of $T'^{(2)}_{j}$ was established in Lemma~\ref{lemma:Stage2}.

\par Define ${\bar T}_1^{(2)} \triangleq \frac{{T}_1^{(2)}}{E_{\text{tot}}}$ and ${\bar T}'^{(2)}_{j} \triangleq \frac{T'^{(2)}_{j}}{E_{\text{tot}}}$. We thus have ${\bar p_{\text{miss}}^{(2)}} = \sum\nolimits_{j=1}^{N-K-1} {\Pr \{ {T}_1^{(2)} <  T'^{(2)}_{j}\} } = \sum\nolimits_{j=1}^{N-K-1} {\Pr \{ {\bar T}_1^{(2)} <  {\bar T}'^{(2)}_{j}\} }$. Since ${\bar T}_1^{(2)}$ and ${\bar T}'^{(2)}_{j}$ are independent, we can proceed by deriving the rate function for each and then combining them to characterize the decay rate of ${\Pr \{ {\bar T}_1^{(2)} <  {\bar T}'^{(2)}_{j}\}}$.

\par Specifically, in similar lines of proof for ${\bar T}_1^{(1)}$, it is standard to show that the rate function of ${\bar T}_1^{(2)}$ is
\begin{align}
I_{{\bar T}_1^{(2)}}(u) = \frac{\Big(\sqrt{u}- \sqrt{\xi_{1}^{(2)}}\Big)^2}{2},~~u\ge 0,
\end{align}
where ${\xi_{1}^{(2)}}= \frac{2|\gamma|^2F_RW_T}{\sigma^2}(\frac{\alpha}{N} + \frac{1-\alpha}{N-K})$.

\par For ${\bar T}'^{(2)}_{j}$, we first evaluate its logarithmic MGF as
\begin{align}
\Lambda_{{{\bar T}'^{(2)}_{j}}}(t_2) &=\mathop {\lim }\limits_{E_{\text{tot}} \to \infty} \frac{1}{{E}_{\text{tot}}}\log{\mathbb E}[ {e^{E_{\text{tot}}t_2{\bar T}'^{(2)}_{j}}}] \\
&=\mathop {\lim }\limits_{E_{\text{tot}} \to \infty} \frac{1}{{E}_{\text{tot}}}\log{{\mathbb E}[ {e^{{T}'^{(2)}_{j}t_2}}]}\\
&=\mathop {\lim }\limits_{E_{\text{tot}} \to \infty} \frac{1}{{E}_{\text{tot}}}\log {\mathbb E}_{T^{(1)}_{(K+j)}}\Big[ {\mathbb E}\big[ e^{t'_2 \frac{E^{(1)}+E^{(2)}}{E^{(2)}}{T}'^{(2)}_{j}}\left|\right. {T^{(1)}_{(K+j)}}\big]\Big] \\
&\mathop = \limits^{(\text{C}.1)}\mathop {\lim }\limits_{E_{\text{tot}} \to \infty} \frac{1}{{E}_{\text{tot}}}\log {\mathbb E}_{T^{(1)}_{(K+j)}}\Bigg[
\frac{e^{\frac{E^{(1)}t'_2}{E^{(2)}(1-2t'_2)}{{T^{(1)}_{(K+j)}}}}}{1 - 2t'_2}\Bigg]  \label{equ:MGF:Stage2:1}\\
&\mathop = \limits^{(\text{C}.1)} \mathop {\lim }\limits_{E_{\text{tot}} \to \infty} \frac{1}{{E}_{\text{tot}}}\log \frac{1}{{1 - 2t'_2}}{\mathbb E}_{{T^{(1)}_{(K+j)}}}\Big[
e^{\frac{E^{(1)}t'_2}{E^{(2)}(1-2t'_2)}{{T^{(1)}_{(K+j)}}}}\Big]\\
&\mathop = \limits^{(\text{C}.1),(\text{C}.2)} \mathop {\lim }\limits_{E_{\text{tot}} \to \infty} \frac{1}{{E}_{\text{tot}}}\Bigg[\log \frac{1}{{1 - 2t'_2}}\prod\limits_{l = 1}^{K + j} {\frac{{\left( {N - l} \right)/2}}{{\left( {N - l} \right)/2 - \frac{E^{(1)}t'_2}{E^{(2)}(1-2t'_2)}}}}\Bigg] \label{equ:MGF:Stage2:2} \\
&\mathop = \limits^{(\text{C}.1),(\text{C}.2)} 0,
\end{align}
where we have
\begin{itemize}
\item auxiliary variable $t'_2 \triangleq t_2\frac{E^{(2)}}{E^{(1)}+E^{(2)}}$;
\item condition $(\text{C}.1)$ reads as
\begin{align}
t'_2 < \frac{1}{2}~~\Rightarrow~~t_2 < \frac{1}{2}\left(1+\frac{E^{(1)}}{E^{(2)}}\right);
\end{align}
\item condition $(\text{C}.2)$ reads as
\begin{align}
&\frac{E^{(1)}t'_2}{E^{(2)}(1-2t'_2)} < \frac{N-{(K+j)}}{2} \\
\Rightarrow~~&t_2 < \frac{1}{2}\left(1+\frac{E^{(1)}}{E^{(2)}}\right)\frac{N-{(K+j)}}{\frac{E^{(1)}}{E^{(2)}} + N-(K+j)},
\end{align}
which is stronger than $(\text{C}.1)$;
\item Equation~\eqref{equ:MGF:Stage2:1} follows from fact~\eqref{equ:conditional:chisquared} of Lemma~\ref{lemma:Stage2} and the MGF of a noncentral chi-squared distributed variable;
\item Equation~\eqref{equ:MGF:Stage2:2} uses the fact that ${T}_{(K+j)}^{(1)}$ (the $(K+j)$th order statistic of $(N-1)$ i.i.d. variables$\sim \chi _2^2 (0)$) is statistically equivalent to $\sum\nolimits_{l = 1}^{(K+j)} {Z_l}$, where $Z_l \sim \text{Exp}(\frac{N-l}{2})$~\cite[Chapter~1]{balakrishnan1998handbook}, and then follows from the MGF of an exponential distributed variable.
\end{itemize}
\par Therefore, the logarithmic MGF of ${\bar T}'^{(2)}_{j}$ exists as an extended real number as
\begin{align}
\Lambda_{{{{\bar T}'^{(2)}_{j}}}}(t_2)= \left\{{\begin{array}{*{20}l}
   0, \text{~~if~}t_2 < \frac{1}{2}\left(1+\frac{E^{(1)}}{E^{(2)}}\right)\frac{N-{(K+j)}}{\frac{E^{(1)}}{E^{(2)}} + N-(K+j)} \triangleq \beta_j\\
   +\infty, \quad \text{otherwise},
\end{array}} \right.
\end{align}
for $j=1, \cdots, N-K-1$. In addition, it is clear that $0 \in {\mathcal D}_{\Lambda}= \{t_2: \Lambda_{{{{\bar T}'^{(2)}_{j}}}}(t_2) < \infty\}$. Therefore, ${\bar T}'^{(2)}_{j}$ satisfies the Gartner-Ellis conditions.
\par The rate function of ${\bar T}'^{(2)}_{j}$ can be calculated as the Fenchel-Legendre transform of $\Lambda_{{{{\bar T}'^{(2)}_{j}}}}(t_2)$:
\begin{align}
I_{{\bar T}'^{(2)}_{j}}(v) &= \mathop {\sup}\limits_{t_2  \in {\mathbb R}} \{ vt_2  - \Lambda(t_2)\} = \mathop {\sup}\limits_{t_2 < \beta_j} \{ vt_2\}= \beta_jv,~~v\ge 0.
\end{align}
\par Since both $\Lambda_{{\bar T}_1^{(2)}}(t_1)$ and $\Lambda_{{{\bar T}'^{(2)}_{j}}}(t_2)$ are essentially smooth, lower semicontinuous functions, the LDP holds for both ${\bar T}_1^{(2)}$ and ${\bar T}'^{(2)}_{j}$ with the good rate function $I_{{\bar T}_1^{(2)}}(u)$ and $I_{{\bar T}'^{(2)}_{j}}(v)$, respectively~\cite[Theorem 2.3.6 (Gartner-Ellis)]{dembo2009large}. We thus further have
\begin{align}
-\mathop {\lim }\limits_{E_{\text{tot}}\to \infty }\frac{1}{{E_{\text{tot}} }} \log {\Pr\{{\bar T}_1^{(2)} < {\bar T}'^{(2)}_{j}\}}&=\mathop {\inf}\limits_{0\le u \le v} \{I_{{\bar T}_1^{(2)}}(u) + I_{{\bar T}'^{(2)}_{j}}(v)\} \\
=&\mathop {\inf}\limits_{0\le u \le v} \Bigg\{\frac{\Big(\sqrt{u}- \sqrt{\xi_{1}^{(2)}}\Big)^2}{2} + \beta_jv\Bigg\} \label{equ:minimization:p2}
\end{align}
\par Using the Karush-Kuhn-Tucker conditions~\cite{boyd2004convex} for the minimization problem in~\eqref{equ:minimization:p2}, it can be shown that the infimum is attained at $u^*= v^* = \frac{\xi_{1}^{(2)}}{(1+ 2\beta_j)^2}$. Therefore, the decay rate of ${\Pr\{{\bar T}_1^{(2)} < {\bar T}'^{(2)}_{j}\}}$ is given by
\begin{align}
-\mathop {\lim }\limits_{E_{\text{tot}}\to \infty }\frac{1}{{E_{\text{tot}} }} \log {\Pr\{{\bar T}_1^{(2)} < {\bar T}'^{(2)}_{j}\}}&= \frac{\Big(\sqrt{u^*}- \sqrt{\xi_{1}^{(2)}}\Big)^2}{2} + \frac{N-K}{2}v^* \nonumber\\
&=\frac{\xi_{1}^{(2)}\beta_j}{1+2\beta_j},~j=1,\cdots, N-K-1.
\end{align}
\par Consequently, the decay rate of ${\bar p_{\text{miss}}^{(2)}} = \sum\nolimits_{j=1}^{N-K-1} {\Pr \{ {\bar T}_1^{(2)} <  {\bar T}'^{(2)}_{j}\} }$ is given by
\begin{align}
-\mathop {\lim }\limits_{E_{\text{tot}}\to \infty }\frac{1}{{E_{\text{tot}} }} \log {\bar p_{\text{miss}}^{(2)}} &= \mathop {\min }\limits_{j \in [1:N - K - 1]} \frac{\xi_{1}^{(2)}\beta_j}{1+2\beta_j}\\
&= \mathop {\min }\limits_{j \in [1:N - K - 1]} \frac{\xi_{1}^{(2)}}{\frac{1}{\beta_j}+2}\\
&= \frac{\xi_{1}^{(2)}}{\frac{1}{\beta_{(N-K-1)}}+2} = \frac{\xi_{1}^{(2)}}{4} \label{equ:decay:rate:stage2:1}
\end{align}
where~\eqref{equ:decay:rate:stage2:1} follows by the fact that
\begin{align}
\frac{1}{\beta_j} &= \frac{2}{1+\frac{E^{(1)}}{E^{(2)}}}\frac{{\frac{E^{(1)}}{E^{(2)}} + N-(K+j)}}{N-{(K+j)}} = \frac{2}{1+ \frac{\alpha}{N}\frac{N-K}{1-\alpha}}\left(1+\frac{\frac{\alpha}{N}\frac{N-K}{1-\alpha}}{N-(K+j)}\right),
\end{align}
which attains its maximum value $2$ when $j = N-K-1$.
\par Finally, taking the minimum of rate functions of ${p_{\text{miss}}^{(1)}}$ and ${\bar p_{\text{miss}}^{(2)}}$ gives a characterization of the rate function of ${\bar p_{\text{miss}}}$. This completes the proof of Proposition~\ref{proposition:LDP}.

\section{Proof of Proposition~\ref{proposition:opt:decayrate}} \label{appendix:proposition:opt:decayrate}
\par Consider the following optimization problem (${\text P}.1$):
\begin{align}
&\max_{\alpha,K} I_{{\bar p_{\text{miss}}}}(\alpha, K) \triangleq \min\left\{ {\frac{{\xi _1^{(1)} }}{{2\left( {1 + \frac{1}{N - K}} \right)}},\frac{{\xi _1^{(2)} }}{4}} \right\}\\
&\text{subject to}~\alpha \in (0,1],~K \in [1:N-1],
\end{align}
where ${\xi _1^{(1)}} = \frac{2|\gamma|^2F_RW_T\alpha}{\sigma^2N}$ and ${\xi _1^{(2)}} = \frac{2|\gamma|^2F_RW_T}{\sigma^2}(\frac{\alpha}{N} + \frac{1-\alpha}{N-K})$.
\par Note that for any given $N$ and some feasible $K$, the first term in $I_{{\bar p_{\text{miss}}}}(\alpha, K)$ monotonically increases as $\alpha$ increases, while the second term monotonically decreases as $\alpha$ increases. Therefore, for any given $K$, $I_{{\bar p_{\text{miss}}}}(\alpha, K)$ attains its maximum when the two terms equal, i.e.,
\begin{align}
\frac{{\xi _1^{(1)} }}{{2\left( {1 + \frac{1}{N - K}} \right)}} =  \frac{{\xi _1^{(2)} }}{4},
\end{align}
which implies that the optimal $\alpha^*$ under a given $K$ (i.e., $\alpha^*(K)$) is
\begin{align}
\alpha^*(K) = \frac{N(N-K+1)}{2(N-K)^2+K(N-K+1)}. \label{equ:alpha:k}
\end{align}
\par As a result, solving $({\text P}.1)$ boils down to finding the optimal $K^*$ for the following maximization problem (${\text P}.2$):
\begin{align}
&\max_{K} I_{{\bar p_{\text{miss}}}}(\alpha^*(K), K) = \frac{2|\gamma|^2F_RW_T}{\sigma^2N}\frac{\alpha^*(K)}{2\big(1+\frac{1}{N-K}\big)} \\
&\quad \quad \quad \quad \quad \quad \quad \quad \quad \propto \frac{\alpha^*(K)}{\big(1+\frac{1}{N-K}\big)} \label{equ:p2:objective}\\
&{\text{subject to}}~K \in [1:N-1].
\end{align}
Specifically, with $\alpha^*(K)$ of~\eqref{equ:alpha:k}, the objective~\eqref{equ:p2:objective} can be rewritten as
\begin{align}
 \frac{N(N-K)}{(N-K)^2+{(N-K)(N-1)}+N}&=\frac{N}{(N-K)+\frac{N}{N-K}+(N-1)}\\
\le&\frac{N}{2\sqrt{N}+(N-1)},
\end{align}
where the equality above attains when $N-K= \sqrt{N}$ in general. Given that $K$ is restricted to an integer in $({\text P}.2)$, then the optimal $K^*$ is
\begin{align}
K^* =  N - \text{round}(\sqrt{N}),
\end{align}
and the optimal $\alpha^*(K^*)$ is evaluated through \eqref{equ:alpha:k} at $K^*$. The corresponding optimal decay rate can be represented as
\begin{align}
I_{{\bar p_{\text{miss}}}}(\alpha^*(K^*), K^*) &= \frac{2|\gamma|^2F_RW_T}{\sigma^2N}\frac{\alpha^*(K^*)}{2\big(1+\frac{1}{N-K^*}\big)} = \frac{|\gamma|^2F_RW_T}{2\sigma^2\Big(N-\frac{K^*(N-K^*-1)}{2(N-K^*)}\Big)}.
\end{align}



\begin{thebibliography}{10}
\providecommand{\url}[1]{#1}
\csname url@samestyle\endcsname
\providecommand{\newblock}{\relax}
\providecommand{\bibinfo}[2]{#2}
\providecommand{\BIBentrySTDinterwordspacing}{\spaceskip=0pt\relax}
\providecommand{\BIBentryALTinterwordstretchfactor}{4}
\providecommand{\BIBentryALTinterwordspacing}{\spaceskip=\fontdimen2\font plus
\BIBentryALTinterwordstretchfactor\fontdimen3\font minus
  \fontdimen4\font\relax}
\providecommand{\BIBforeignlanguage}[2]{{%
\expandafter\ifx\csname l@#1\endcsname\relax
\typeout{** WARNING: IEEEtran.bst: No hyphenation pattern has been}%
\typeout{** loaded for the language `#1'. Using the pattern for}%
\typeout{** the default language instead.}%
\else
\language=\csname l@#1\endcsname
\fi
#2}}
\providecommand{\BIBdecl}{\relax}
\BIBdecl

\bibitem{min2019ICC}
M.~Li, C.~Liu, S.~V. Hanly, I.~B. Collings, and P.~Whiting, ``Beam alignment
  with two-stage search for millimeter-wave communications,'' 2019, in
  \textit{Proc. IEEE ICC}, Shanghai, China.

\bibitem{pi2011introduction}
Z.~Pi and F.~Khan, ``An introduction to millimeter-wave mobile broadband
  systems,'' \emph{{IEEE} Commun. Mag.}, vol.~49, no.~6, 2011.

\bibitem{andrews2014will}
J.~G. Andrews, S.~Buzzi, W.~Choi, S.~V. Hanly, A.~Lozano, A.~C. Soong, and
  J.~C. Zhang, ``What will 5{G} be?'' \emph{IEEE J. Sel. Areas Comms.},
  vol.~32, no.~6, pp. 1065--1082, 2014.

\bibitem{xiao2017millimeter}
M.~Xiao, S.~Mumtaz, Y.~Huang, L.~Dai, Y.~Li, M.~Matthaiou, G.~K. Karagiannidis,
  E.~Bj{\"o}rnson, K.~Yang, I.~Chih-Lin \emph{et~al.}, ``Millimeter wave
  communications for future mobile networks,'' \emph{{IEEE} J. Sel. Areas
  Commun.}, vol.~35, no.~9, pp. 1909--1935, 2017.

\bibitem{lee2018spectrum}
J.~Lee~\textit{et al.}, ``Spectrum for {5G}: Global status, challenges, and
  enabling technologies,'' \emph{{IEEE} Commun. Mag.}, vol.~56, no.~3, pp.
  12--18, 2018.

\bibitem{choi2016millimeter}
J.~Choi, V.~Va, N.~Gonzalez-Prelcic, R.~Daniels, C.~R. Bhat, and R.~W. Heath,
  ``Millimeter-wave vehicular communication to support massive automotive
  sensing,'' \emph{{IEEE} Commun. Mag.}, vol.~54, no.~12, pp. 160--167, 2016.

\bibitem{liu2018Mag}
C.~Liu, M.~Li, S.~V. Hanly, P.~Whiting, and I.~B. Collings, ``Millimeter wave
  small cells: Base station discovery, beam alignment and system design
  challenges,'' \emph{{IEEE} Wireless Commun. Mag.}, vol.~25, no.~4, pp.
  40--46, August 2018.

\bibitem{wang2009beam}
J.~Wang, Z.~Lan, C.-W. Pyo, T.~Baykas, C.-S. Sum, M.~A. Rahman, J.~Gao,
  R.~Funada, F.~Kojima, H.~Harada \emph{et~al.}, ``Beam codebook based
  beamforming protocol for multi-{G}bps millimeter-wave {WPAN} systems,''
  \emph{{IEEE} J. Sel. Areas Commun.}, vol.~27, no.~8, pp. 1390--1399, 2009.

\bibitem{hur2013millimeter}
S.~Hur, T.~Kim, D.~J. Love, J.~V. Krogmeier, T.~A. Thomas, and A.~Ghosh,
  ``Millimeter wave beamforming for wireless backhaul and access in small cell
  networks,'' \emph{{IEEE} Trans. Commun.}, vol.~61, no.~10, pp. 4391--4403,
  2013.

\bibitem{alkhateeb2014channel}
A.~Alkhateeb, O.~El~Ayach, G.~Leus, and R.~W. Heath, ``Channel estimation and
  hybrid precoding for millimeter wave cellular systems,'' \emph{IEEE J. Sel.
  Sig. Processing}, vol.~8, no.~5, pp. 831--846, 2014.

\bibitem{Xiao2016}
Z.~Xiao, T.~He, P.~Xia, and X.~G. Xia, ``Hierarchical codebook design for
  beamforming training in millimeter-wave communication,'' \emph{{IEEE} Trans.
  Wireless Commun.}, vol.~15, no.~5, pp. 3380--3392, 2016.

\bibitem{liu2017Jsac}
C.~Liu, M.~Li, S.~V. Hanly, I.~B. Collings, and P.~Whiting, ``Millimeter wave
  beam alignment: Large deviations analysis and design insights,'' \emph{{IEEE}
  J. Sel. Areas Commun.}, vol.~35, no.~7, pp. 1619--1631, July 2017.

\bibitem{zhang2017codebook}
J.~Zhang, Y.~Huang, Q.~Shi, J.~Wang, and L.~Yang, ``Codebook design for beam
  alignment in millimeter wave communication systems,'' \emph{IEEE Transactions
  on Communications}, vol.~65, no.~11, pp. 4980--4995, 2017.

\bibitem{haghighatshoar2016beam}
S.~Haghighatshoar and G.~Caire, ``The beam alignment problem in mmwave wireless
  networks,'' in \emph{Proc. 50th Asilomar Conf. on Signals, Syst. and
  Comput.}, 2016, pp. 741--745.

\bibitem{kokshoorn2017millimeter}
M.~Kokshoorn, H.~Chen, P.~Wang, Y.~Li, and B.~Vucetic, ``Millimeter wave {MIMO}
  channel estimation using overlapped beam patterns and rate adaptation,''
  \emph{{IEEE} Trans. Signal Process.}, vol.~65, no.~3, pp. 601--616, 2017.

\bibitem{alkhateeb2017initial}
A.~Alkhateeb, Y.-H. Nam, M.~S. Rahman, J.~Zhang, and R.~W. Heath, ``Initial
  beam association in millimeter wave cellular systems: Analysis and design
  insights,'' \emph{{IEEE} Trans. Wireless Commun.}, vol.~16, no.~5, pp.
  2807--2821, 2017.

\bibitem{va2018inverse}
V.~Va, J.~Choi, T.~Shimizu, G.~Bansal, and R.~W. Heath, ``Inverse multipath
  fingerprinting for millimeter wave {V2I} beam alignment,'' \emph{{IEEE}
  Trans. Veh. Technol.}, vol.~67, no.~5, pp. 4042--4058, 2018.

\bibitem{3GPPR1}
3GPP-R1-167672, ``Synchronization in {NR} considering beam sweeping,'' 2016.

\bibitem{3GPPR2}
3GPP-R1-167543, ``Beam management considerations for above 6 {GHz} {NR},''
  2016.

\bibitem{raghavan2018steady}
V.~Raghavan, J.~Luo, R.~N. Challa, Q.~Liu, J.~Li, A.~Y. Gorokhov, and
  A.~Touboul, ``Steady-state beam scanning and codebook generation,'' 2018,
  {US} Patent App. 15/684,861.

\bibitem{xiao2017codebook}
Z.~Xiao, P.~Xia, and X.-G. Xia, ``Codebook design for millimeter-wave channel
  estimation with hybrid precoding structure,'' \emph{IEEE Trans. Wireless
  Comms.}, vol.~16, no.~1, pp. 141--153, 2017.

\bibitem{xiao2018enhanced}
Z.~Xiao, H.~Dong, L.~Bai, P.~Xia, and X.-G. Xia, ``Enhanced channel estimation
  and codebook design for millimeter-wave communication,'' \emph{{IEEE} Trans.
  Veh. Technol.}, vol.~67, no.~10, pp. 9393--9405, 2018.

\bibitem{lin2017subarray}
C.~Lin, G.~Y. Li, and L.~Wang, ``Subarray-based coordinated beamforming
  training for {mmWave} and sub-{THz} communications,'' \emph{{IEEE} J. Sel.
  Areas Commun.}, vol.~35, no.~9, pp. 2115--2126, 2017.

\bibitem{liu2017design}
C.~Liu, M.~Li, I.~B. Collings, S.~V. Hanly, and P.~Whiting, ``Design and
  analysis of transmit beamforming for millimeter wave base station
  discovery,'' \emph{{IEEE} Trans. Wireless Commun.}, vol.~16, no.~2, pp.
  797--811, 2017.

\bibitem{barati2014dreictional}
C.~Barati~Nt., S.~Hosseini, S.~Rangan, P.~Liu, T.~Korakis, S.~Panwar, and
  T.~Rappaport, ``Directional cell discovery in millimeter wave cellular
  networks,'' \emph{IEEE Trans. Wireless Comms.}, vol.~14, no.~12, pp.
  6664--6678, 2015.

\bibitem{Castellanos2018ICC}
R.~M. Castellanos \emph{et~al.}, ``Hybrid multi-user precoding with amplitude
  and phase control,'' in \emph{Proc. IEEE ICC}, 2018, pp. 1--6.

\bibitem{karacora2019hybrid}
Y.~Karacora, A.~Kariminezhad, and A.~Sezgin, ``Hybrid beamforming: Where should
  the analog power amplifiers be placed?'' \emph{arXiv preprint
  arXiv:1902.09511}, 2019.

\bibitem{zhang2014achieving}
E.~Zhang and C.~Huang, ``On achieving optimal rate of digital precoder by
  rf-baseband codesign for mimo systems,'' in \emph{Proc. IEEE 80th Veh.
  Technol. Conf.}, 2014, pp. 1--5.

\bibitem{IBM_beamswitch}
A.~Valdes-Garcia, ``Millimeter-wave phased array for 5g,''
  http://www.winlab.rutgers.edu/iab/2018-01/Slides/04\%20-\%20IBM\_5G\_PAWR\_June\_1st.pdf.

\bibitem{constantine2005antenna}
A.~Constantine~Balanies, \emph{Antenna theory analysis and design}.\hskip 1em
  plus 0.5em minus 0.4em\relax John Wiley \& Sons Inc., New York, 2005.

\bibitem{balakrishnan1998handbook}
N.~Balakrishnan and C.~R. Rao, \emph{Handbook of statistics. v.16: Order
  statistics: theory and methods}.\hskip 1em plus 0.5em minus 0.4em\relax North
  Holland, Amsterdam (Netherlands), 1998.

\bibitem{gradshteyn2014table}
I.~S. Gradshteyn and I.~M. Ryzhik, \emph{Table of integrals, series, and
  products}.\hskip 1em plus 0.5em minus 0.4em\relax Academic press, 2014.

\bibitem{muhi2010modelling}
Z.~Muhi-Eldeen, L.~P. Ivrissimtzis, and M.~Al-Nuaimi, ``Modelling and
  measurements of millimetre wavelength propagation in urban environments,''
  \emph{IET microwaves, antennas and propagation}, vol.~4, no.~9, pp.
  1300--1309, 2010.

\bibitem{samimi201628}
M.~K. Samimi, G.~R. MacCartney, S.~Sun, and T.~S. Rappaport, ``28 ghz
  millimeter-wave ultrawideband small-scale fading models in wireless
  channels,'' in \emph{Proc. IEEE 83rd Veh. Technol. Conf.}, 2016, pp. 1--6.

\bibitem{Akdeniz2014}
M.~Akdeniz, Y.~Liu, M.~Samimi, S.~Sun, S.~Rangan, T.~Rappaport, and E.~Erkip,
  ``Millimeter wave channel modeling and cellular capacity evaluation,''
  \emph{IEEE J. Sel. Areas Comms.}, vol.~32, no.~6, pp. 1164--1179, June 2014.

\bibitem{fan2018flat}
W.~Fan, C.~Zhang, and Y.~Huang, ``Flat beam design for massive mimo systems via
  riemannian optimization,'' \emph{{IEEE} Wireless Commun. Lett.}, 2018,
  available via {Early} {Access}.

\bibitem{cui2012two}
G.~Cui, L.~Kong, X.~Yang, and D.~Ran, ``Two useful integrals involving
  generalised marcum {Q}-function,'' \emph{Electronics letters}, vol.~48,
  no.~16, pp. 1017--1018, 2012.

\bibitem{dembo2009large}
A.~Dembo and O.~Zeitouni, \emph{Large deviations techniques and
  applications}.\hskip 1em plus 0.5em minus 0.4em\relax Berlin, Germany:
  Springer, 2009.

\bibitem{boyd2004convex}
S.~Boyd and L.~Vandenberghe, \emph{Convex optimization}.\hskip 1em plus 0.5em
  minus 0.4em\relax Cambridge, U.K.: Cambridge Univ. Press, 2004.

\end{thebibliography}
\end{document}